\documentclass[12pt]{revtex4}

\begin{document}

\title{Theory of direct simulation Monte Carlo method}

\author{Hasan Karabulut}

\address {Rize University,\\
Faculty of Arts and Sciences, Physics department,\\
53100 Rize/TURKEY}

\author{Huriye Ar\i man Karabulut}

\address {Rize University,\\
Faculty of fisheries,\\
53100 Rize/TURKEY}

\begin{abstract}
A treatment of direct simulation Monte Carlo method (DSMC) as a
Markov process with a master equation is given and the corresponding
master equation is derived. A hierarchy of equations for the reduced
probability distributions is derived from the master equation. An
equation similar to the Boltzmann equation for single particle
probability distribution is derived using assumption of molecular
chaos. It is shown that starting from an uncorrelated state, the
system remains uncorrelated always in the limit $N\rightarrow \infty
,$ where $N$ is the number of particles. Simple applications of the
formalism to direct simulation money games are given as examples to
the formalism. The formalism is applied to the direct simulation of
homogenous gases. It is shown that appropriately normalized single
particle probability distribution satisfies the Boltzmann equation
for simple gases and Wang Chang-Uhlenbeck equation for a mixture of
molecular gases. As a consequence of this development we derive
Birds no time counter algorithm. We extend the analysis to the
inhomogenous gases and define a new direct simulation algorithm for
this case. We show that single particle probability
distribution satisfies the Boltzmann equation in our algorithm in the limit $%
N\rightarrow \infty ,$ $V_{k}\rightarrow 0,$ $\Delta t\rightarrow 0$ where $%
V_{k}$ is the volume $k^{th}$ cell. We also show that that our
algorithm and Bird's algorithm approach each other in the limit
$N_{k}\rightarrow \infty$ where $N_{k}$ is the number of particles
in the volume $V_{k}$.
\end{abstract}

\maketitle 

\section{Introduction}

Direct simulation Monte Carlo method (DSMC)\cite{Bird94} is a
standard method to solve the Boltzmann equation numerically. In this method
one divides space into cells of volume $V_{k}$ $(k=1,2,3,...)$ and takes a
large number ($N$) of simulated particles ($10^{3}-10^{6}$) to represent
real gas molecules. The time evolution of the gas for a short time period $%
\Delta t$ is calculated in two steps. In the first step some pairs of
particles in the same cell are chosen randomly and are allowed to collide
without changing their positions. A collision is allowed with a probability
proportional to $u\Sigma $ where $u$ is the relative velocity and $\Sigma $
is the total cross section. In the second step all particles are propagated
without collisions for a time $\Delta t$.

The method is invented by Bird and Bird introduced the method based on
physical arguments. A seminal paper of Bird\cite{Bird70} gives somewhat
heuristic arguments to justify its use to solve the Boltzmann equation. One
variant of the method was derived by Nanbu\cite{Nanbu80} starting from the
Boltzmann equation. Also it appears that essentially the same stochastic
algorithms for a homogenous gas were invented independently by people
interested in using them as a pedagogical tool to demonstrate evolution of a
gas toward Maxwell-Boltzmann(MB) distribution.\cite{Novak70} \cite{Eger82}
\cite{Bonomo84}. In order to represent time evolution of the real gas such
methods should converge to the true solution of the Boltzmann equation in
the limit of $N\rightarrow \infty ,$ $V_{k}\rightarrow 0,$ $\Delta
t\rightarrow 0.$ Convergence proofs were given by Babovsky\cite{Babovsky1}
and Babovsky and Illner\cite{Babovsky2} for Nanbu's method and by Wagner\cite
{Wagner92} for Bird's method.

The cited convergence proofs are very formal and they appear to be written
for mathematicians. In this paper we give a simple derivation of Birds no
time counter algorithm. We also show that, in DSMC, appropriately normalized
single particle probability distribution satisfies Boltzmann equation for
simple gases and Wang Chang-Uhlenbeck equation for molecular gases and their
mixtures. The language of this development is familiar to the physicist from
the well known BBGKY hierarchy.

In the next section we develop a general formalism for direct simulation. In
order to demonstrate usefulness of the formalism we apply it to some simple
money games. In the third section we apply the formalism to homogenous gases
and show that, if appropriate collision kernels are chosen, the one particle
probability distribution obeys the Boltzmann equation for simple gases and
the Wang Chang-Uhlenbeck equation for molecular gases and their mixtures. In
the fourth section we derive DSMC\ algorithm for inhomogeneous gases.
Finally in the last section we give a summary and discussion.

\section{Direct simulation as a Markov process}

\subsection{The Master Equation}

Assume that we have an assembly of things we call 'particles'. Particles can
be real particles in a gas or humans or anything you can imagine. There are $%
N$ particles in the assembly where $N$ is a very large number. Each member
of the assembly can be in any one of the 'states' where states are labeled
by the parameter $\mu $. For a real gas $\mu $ can be velocity vectors and
for an assembly of people $\mu $ can be the money in their pocket on bank
account. The $\mu $ can be discrete or continuous and it can stand for a
collection of indices that can be both continuous and discrete. For the rest
of this section we will treat $\mu $ as a continuous index. Integration over
$\mu $ is actually integration over the continuous indices and summation
over the discrete indices that $\mu $ stands for.

We play a stochastic game with this assembly. We randomly pick pairs of
particles and force them to 'collide'. A collision is an event that the
particles change their states with a prescribed probability. Suppose we
picked particles with states $\mu _{A}$ and $\mu _{B}.$ The probability that
they will end up with state labels $\mu _{C}$ and $\mu _{D}$ in the volume $%
d\mu _{C}d\mu _{D}$ is $T(\mu _{A},\mu _{B};\mu _{C},\mu _{D})d\mu _{C}d\mu
_{D}$ where $T(\mu _{A},\mu _{B};\mu _{C},\mu _{D})$ is the collision
kernel. Collision kernel is assumed to be symmetric
\begin{eqnarray}
T(\mu _{A},\mu _{B};\mu _{C},\mu _{D}) &=&T(\mu _{C},\mu _{D};\mu _{A},\mu
_{B}),  \label{a10} \\
T(\mu _{A},\mu _{B};\mu _{C},\mu _{D}) &=&T(\mu _{B},\mu _{A};\mu _{D},\mu
_{C}).  \label{a20}
\end{eqnarray}
Also the probabilities are normalized
\begin{equation}
\int T(\mu _{A},\mu _{B};\mu _{C},\mu _{D})\,d\mu _{C}\,d\mu _{D}=\int T(\mu
_{A},\mu _{B};\mu _{C},\mu _{D})\,d\mu _{A}\,d\mu _{B}=1.  \label{a30}
\end{equation}

We define N-particle probability distribution $f^{(N)}(\mu _{1},\mu
_{2},...,\mu _{N};n)$ such that $f^{(N)}(\mu _{1},\mu _{2},...,\mu
_{N};n)d\mu _{1}d\mu _{2},...,d\mu _{N}$ is the probability of finding the
particles $1,2,...,N$ in the $d\mu _{1}d\mu _{2},...,d\mu _{N}$ phase space
volume after the $n^{th}$ collision. Since the particles are identical the $%
f^{(N)}(\mu _{1},\mu _{2},...,\mu _{N};n)$ is assumed to be completely
symmetric
\begin{equation}
f^{(N)}(\mu _{1},...,\mu _{j},...,\mu _{i},...,\mu _{N};n)=f^{(N)}(\mu
_{1},...,\mu _{i},...,\mu _{j},...,\mu _{N};n).  \label{a40}
\end{equation}

We define reduced M-particle distribution as
\begin{equation}
f^{(M)}(\mu _{1},...,\mu _{N};n)=\int f^{(N)}(\mu _{1},...,\mu _{N};n)\,d\mu
_{M+1}\,d\mu _{M+2},...,d\mu _{N}.  \label{a44}
\end{equation}
We will denote $f^{(M)}(\mu _{1},....,\mu _{M};n)$ $(M=1,2,...,N)$ as $%
f^{(M)}(\mu ;n)$ shortly. As a convenient notation we also define $%
f_{ij}^{(M)}(\mu _{A},\mu _{B};n)$ as
\begin{equation}
f_{ij}^{(M)}(\mu _{A},\mu _{B};n)=f^{(M)}(\mu _{1},...,\mu _{i}=\mu
_{A},...,\mu _{j=}\mu _{B},...,\mu _{M};n),  \label{a45}
\end{equation}
where $\mu _{i}$ and $\mu _{j}$ are replaced with $\mu _{A}$ and $\mu _{B}$
in $f^{(M)}(\mu _{1},...,\mu _{M};n)$. Examples are
\begin{eqnarray}
f_{31}^{(N)}(\mu _{A},\mu _{B};n) &=&f(\mu _{B},\mu _{2},\mu _{A},\mu
_{4},...,\mu _{N};n)  \label{a48} \\
f_{24}^{(N)}(\mu _{A},\mu _{B};n) &=&f(\mu _{1},\mu _{A},\mu _{3},\mu
_{B},\mu _{5},...,\mu _{N};n)  \label{a49}
\end{eqnarray}

We are ready to start now. The equation satisfied by the $f^{(N)}(\mu ;n)$
is given by
\begin{equation}
f^{(N)}(\mu ;n+1)=\frac{1}{N(N-1)}\sum_{i=1}^{N}\sum_{j\neq i}^{N}\int
f_{ij}^{(N)}(\mu _{A},\mu _{B};n)\,T(\mu _{A},\mu _{B};\mu _{i},\mu
_{j})\,d\mu _{A}\,d\mu _{B}.  \label{a66}
\end{equation}
The meaning of this equation is clear. If the last pair we collided is $i,j$
molecules, the probability of having $\mu _{i},\mu _{j}$ pairs at the end of
collision is the probability of having initial states $\mu _{A},\mu _{B}$
(represented by $f_{ij}^{(N)}(\mu _{A},\mu _{B};n)d\mu _{A}d\mu _{B}$)
multiplied by the probability of ending with $\mu _{i},\mu _{j}$
(represented by $T(\mu _{A},\mu _{B};\mu _{i},\mu _{j})$). The sum over $i,j$
and the factor $1/N(N-1)$ takes care of the fact that all pairs (respecting
order of the molecules) are possible with the probability $1/N(N-1).$ The
state of the system after $n+1$ collisions depends on the state of system
after $n$ collisions and the direct simulation game is a Markov process
actually. The eq.(\ref{a66}) is the master equation for this stochastic
process.

In order to see clearly how this equation is derived let us multiply this
with $d\mu _{1}d\mu _{2}...d\mu _{N}$. The left hand side is
\begin{equation}
f^{(N)}(\mu ;n+1)d\mu _{1}d\mu _{2}...d\mu _{N}  \label{ad10}
\end{equation}
and it is the probability of the system being in the phase space volume $%
d\mu _{1}d\mu _{2}...d\mu _{N}$ after the $(n+1)^{th}$ collision. On the
right side we have
\begin{equation}
\frac{1}{N(N-1)}\sum_{i=1}^{N}\sum_{j\neq i}^{N}\int f_{ij}^{(N)}(\mu
_{A},\mu _{B};n)T(\mu _{A},\mu _{B};\mu _{i},\mu _{j})d\mu _{A}d\mu _{B}d\mu
_{1}d\mu _{2}...d\mu _{N}.  \label{ad20}
\end{equation}
(Here the integration is over $\mu _{A}$ and $\mu _{B}$ only) In order to
interpret this lets us look at $i=1$ and $j=2$ term. It is the following
term
\begin{eqnarray}
&&\left[ \frac{1}{N(N-1)}\right] \left[ f^{(N)}(\mu _{A},\mu _{B},\mu
_{3},\mu _{4,}...,\mu _{N})d\mu _{A}d\mu _{B}d\mu _{3}d\mu _{4}...d\mu
_{N}\right]  \nonumber \\
&&\times \left[ T(\mu _{A},\mu _{B};\mu _{1},\mu _{2})d\mu _{1}d\mu
_{2}\right]  \label{ad30}
\end{eqnarray}
integrated over $\mu _{A},$ $\mu _{B}.$ In this form the terms under the
integration are product of three probabilities. $1/N(N-1)$ is the
probability of choosing $i=1,j=2\,$pair. The second parenthesis is the
probability of finding the system in $d\mu _{A}d\mu _{B}d\mu _{3}d\mu
_{4}...d\mu _{N}$ phase space volume before the collision. The last
parenthesis is the probability of taking particles one and two from $d\mu
_{A}d\mu _{B}$ to $d\mu _{1}d\mu _{2}$ interval after the collision. When
integrated over $\mu _{A},$ $\mu _{B}$ this term becomes the probability of
arriving in $d\mu _{1}d\mu _{2}...d\mu _{N}$ phase space volume after $%
(n+1)^{th}$ collision via a collision between particles one and two. If all
such term are summed over $i$ and $j$ we find the probability of probability
of arriving in $d\mu _{1}d\mu _{2}...d\mu _{N}$ phase space volume after $%
(n+1)^{th}$ collision which is the same as eq.(\ref{ad10}).

\subsection{Asymptotic Behavior of the Master Equation}

Let us introduce a short notation for state variables:
\begin{equation}
\begin{array}{ll}
X=(x_{1},x_{2,}...,x_{N}) & dX=dx_{1}dx_{2}...dx_{N} \\
Y=(y_{1},y_{2,}...,y_{N}) & dY=dy_{1}dy_{2}...dy_{N} \\
Z=(z_{1},z_{2,}...,z_{N}) & dZ=dz_{1}dz_{2}...dz_{N}
\end{array}
.  \label{ab10}
\end{equation}
Then the Master equation can be written in the form
\begin{equation}
f(X;n+1)=\int P(X,Y)f(Y;n)dY.  \label{ab20}
\end{equation}
The $P(X,Y)$ has $N(N-1)$ terms and each one of the terms contains $N-2$
delta functions. For example $i=1$, $j=2$ term reads as
\begin{equation}
\frac{1}{N(N-1)}T(x_{1},x_{2};y_{1},y_{2})\delta (x_{3}-y_{3})...\delta
(x_{N}-y_{N}).  \label{ab30}
\end{equation}
The general expression for $P(X,Y)$ is
\begin{equation}
P(X,Y)=\frac{1}{N(N-1)}\sum_{i=1}^{N}\sum_{j\neq i}^{N}\left(
T(x_{i},x_{j};y_{i},y_{j})\prod_{k\neq i,j}^{N}\delta (x_{k}-y_{k})\right)
\label{ab40}
\end{equation}
The $P(X,Y)dX$ is the probability that the system jumps from $Y$ to $dX$
phase space volume after a collision. As can be seen directly from eq.(\ref
{ab40}) it is also symmetric: $P(X,Y)=P(Y,X).$ As a probability density it
satisfies the normalization condition
\begin{equation}
\int P(X,Y)dX=\int P(X,Y)dY=1.  \label{ab50}
\end{equation}
We will need convolution of $P(X,Y)$ shortly. Let us define $W(X,Y)$ as
\begin{equation}
W(X,Y)=\int P(X,Z)P(Y,Z)dZ  \label{ab60}
\end{equation}
It is easily seen that $W(X,Y)$ is symmetric ($W(X,Y)=W(Y,X)$) and it also
satisfies a normalization condition
\begin{equation}
\int W(X,Y)dX=\int W(X,Y)dY=1.  \label{ab65}
\end{equation}

Now we are ready to discuss asymptotic behavior or the master equation. Let
us form $\int f^{2}(X;n+1)dX$ as
\begin{eqnarray}
\int f^{2}(X;n+1)dX &=&\int dX\left( \int P(X,Y)f(Y;n)dY\right) \left( \int
P(X,Z)f(Z;n)dZ\right)  \label{ab70} \\
&=&\int W(Y,Z)f(Y;n)f(Z;n)dYdZ  \label{ab80}
\end{eqnarray}
We can also write $\int f^{2}(X;n)dX$ as
\begin{equation}
\int f^{2}(X;n)dX=\int W(Y,Z)f^{2}(Y)dYdZ=\int W(Y,Z)f^{2}(Z)dYdZ
\label{ab90}
\end{equation}
which follows from eq.(\ref{ab65}). Using these two relations we can write
the following
\begin{eqnarray}
\int f^{2}(X;n+1)dX-\int f^{2}(X;n)dX &=&\int W(Y,Z)f(Y;n)f(Z;n)dYdZ
\label{ab100} \\
&&-\frac{1}{2}\int W(Y,Z)f^{2}(Y)dYdZ  \nonumber \\
&&-\frac{1}{2}\int W(Y,Z)f^{2}(Z)dYdZ  \nonumber
\end{eqnarray}
The right side can be written as
\begin{equation}
\int f^{2}(X;n+1)dX-\int f^{2}(X;n)dX=-\frac{1}{2}\int W(Y,Z)\left(
f(Y;n)-f(Z;n)\right) ^{2}dYdZ.  \label{ab110}
\end{equation}
Since $W(Y,Z)$ is always nonnegative the expression on the right is always
negative or zero. This means $\int f^{2}(X;n)dX$ decreases after each
collision. The decrease stops when $f(Y;n)-f(Z;n)=0$ for all $Y$ and $Z$ and
this means $f(X;n)$ must be a constant. The equilibrium is reached when $%
f(X;n)$ is microcanonical distribution.

There is a final point to be discussed here. The above argument proves that
the probability density in the direct simulation always converges towards
microcanonical distribution. If the phase space is divided in separate
regions such that collisions cannot take the system from one region to
another then the above argument must be modified. If $Y$ and $Z$ belong to
different regions then $W(Y,Z)=0$ and $f(Y;n)-f(Z;n)=0$ is not required. But
if $Y$ and $Z$ belong to the same region then $W(Y,Z)\neq 0$ and $%
f(Y;n)-f(Z;n)=0$ is required. This means that $f(X;n)$ must be a constant in
each region asymptotically but they can be different constants. For direct
simulation of a gas total energy and total momentum are conserved and the
system stays on a constant total energy-total momentum shell. Asymptotically
the $f(X;n)$ will be constant on each shell but they will be different
constant for different shells.

\subsection{The hierarchy of Reduced probability distributions}

If we integrate the master equation over $d\mu _{M+1},\mu _{M+2},...,\mu
_{N} $ we obtain the equation
\begin{eqnarray}
f^{(M)}(\mu \mathbf{;}n+1) &=&\frac{(N-M)(N-M-1)}{N(N-1)}\,\,f^{(M)}(\mu
\mathbf{;}n)  \label{a60} \\
&&+\frac{2(N-M)}{N(N-1)}\sum_{i=1}^{M}\int f_{i,M+1}^{(M+1)}(\mu _{A},\mu
_{B};n)\,T(\mu _{A},\mu _{B};\mu _{i},\mu _{C})\,d\mu _{A}\,d\mu _{B}\,d\mu
_{C}  \nonumber \\
&&+\frac{M(M-1)}{N(N-1)}\sum_{i=1}^{M}\sum_{j\neq i}^{M}\int
f_{i,j}^{(M)}(\mu _{A},\mu _{B};n)\,T(\mu _{A},\mu _{B};\mu _{i},\mu
_{j})\,d\mu _{A}\,d\mu _{B}.  \nonumber
\end{eqnarray}
The $f^{(M)}(\mu \mathbf{;}n+1)$ depends on $f^{(M+1)}(\mu ;n)$ and this
represents a hierarchy of equations similar to the well-known BBGKY hierarchy%
\cite{Huang}.

The first equation in the hierarchy is
\begin{eqnarray}
f^{(1)}(\mu \mathbf{;}n+1) &=&(1-2/N)\,f^{(1)}(\mu \mathbf{;}n)  \label{a70}
\\
&&+\frac{2}{N}\int f^{(2)}(\mu _{A},\mu _{B};n)\,T(\mu _{A},\mu _{B};\mu
_{C},\mu )\,d\mu _{A}\,d\mu _{B}\,d\mu _{C}.  \nonumber
\end{eqnarray}
If we make the assumption of molecular chaos (AMC)
\begin{equation}
f^{(2)}(\mu _{A},\mu _{B};n)=f^{(1)}(\mu _{A};n)\,f^{(1)}(\mu _{B};n),
\label{a80}
\end{equation}
we obtain a nonlinear equation for $f^{(1)}(\mu ;n)$ similar to the
Boltzmann equation.

From now on we will suppress the superscript $(1)$ in $f^{(1)}(\mu \mathbf{;}%
\tau )$ wherever it does not cause confusion. Using the relation
\begin{equation}
f(\mu ,n)=\int f(\mu ,n)\,f(\mu _{C},n)\,T(\mu _{A},\mu _{B};\mu _{C},\mu
)\,d\mu _{A}\,d\mu _{B}\,d\mu _{C},  \label{a90}
\end{equation}
which follows from Eq.(\ref{a30}) and the normalization of $f(\mu _{C})$ and
imposing the assumption of molecular chaos we can write eq.(\ref{a70}) as
\begin{equation}
f(\mu \mathbf{;}n+1)=f(\mu \mathbf{;}n)+\frac{2}{N}\int [f,f]\,T(\mu
_{A},\mu _{B};\mu _{C},\mu )\,d\mu _{A}\,d\mu _{B}\,d\mu _{C}  \label{a101}
\end{equation}
\begin{equation}
\lbrack f,f]=f(\mu _{A},n)\,f(\mu _{B},n)-f(\mu _{C},n)\,f(\mu ,n)
\label{a103}
\end{equation}

A\ second simplification occurs for large $N.$ The $2/N$ appearing in eq.(%
\ref{a101}) is a small number and we can take $\tau =2n/N$ as a continuous
parameter which we call the collision time. Then $\Delta \tau =2/N$ and $%
\left[ f(\mu \mathbf{;}n+1)-f(\mu \mathbf{;}n)\right] /\Delta \tau $ can be
taken as $\partial f(\mu \mathbf{,}\tau )/\partial \tau $. The eq.(\ref{a101}%
) can be written in either of the following forms:
\begin{eqnarray}
\frac{\partial f(\mu ,\tau \mathbf{)}}{\partial \tau } &=&\int [f,f]\,T(\mu
_{A},\mu _{B};\mu _{C},\mu )\,d\mu _{A}\,d\mu _{B}\,d\mu _{C}.  \label{a130}
\\
\frac{\partial f(\mu ,\tau \mathbf{)}}{\partial \tau } &=&-f(\mu )+\int
f(\mu _{A})f\,(\mu _{B})T(\mu _{A},\mu _{B};\mu _{C},\mu )\,d\mu _{A}\,d\mu
_{B}\,d\mu _{C}.  \label{a131}
\end{eqnarray}

We will call the first equation in the hierarchy 'the first equation'
briefly for the rest of the paper. In latter parts of this paper we will
call the integral on the right side of eq.(\ref{a130}) 'the collision
integral'. From now on we will also suppress the collision time $\tau $ in $%
f(\mu \mathbf{,}\tau )$ wherever it is convenient$.$

\subsection{Justification of assumption of molecular chaos}

The only thing in this paper that is not fully rigorous is the assumption of
molecular chaos. In order to have assumption of molecular chaos valid from
the beginning we must start from an uncorrelated state
\begin{equation}
f^{(N)}(\mu _{1},\mu _{2},...,\mu _{N};n=0)=h(\mu _{1})\,h(\mu
_{2})....h(\mu _{N}),  \label{a83}
\end{equation}
which is what is done in direct simulations mostly. The master equation eq.(%
\ref{a66}) should be used to justify AMC. For finite $N$ the AMC\ is not
strictly valid and the AMC\ should get better and better as $N\rightarrow
\infty $. For $M/N<<1$ the eq. (\ref{a60}) is written as
\begin{eqnarray}
f^{(M)}(\mu \mathbf{;}n+1) &=&(1-2M/N)\,\,f^{(M)}(\mu \mathbf{;}n)+O(1/N^{2})
\label{a133} \\
&&+\frac{2}{N}\sum_{i=1}^{M}\int f_{i,M+1}^{(M+1)}(\mu _{A},\mu _{B};n)
\nonumber \\
&&\times \,T(\mu _{A},\mu _{B};\mu _{i},\mu _{C})\,d\mu _{A}\,d\mu
_{B}\,d\mu _{C}  \nonumber
\end{eqnarray}
where $O(1/N^{2})$ are the terms of order $1/N^{2}$. If we invoke collision
time $\tau =2n/N$ again and write $\left[ f^{(M)}(\mu \mathbf{;}%
n+1)-f^{(M)}(\mu \mathbf{;}n)\right] /\Delta \tau =\partial f^{(M)}(\mu
\mathbf{;}\tau )/\partial \tau $ and we take the limit $N\rightarrow \infty $
we obtain
\begin{eqnarray}
\frac{\partial f^{(M)}(\mu \mathbf{;}\tau )}{\partial \tau }
&=&-Mf^{(M)}(\mu \mathbf{;}\tau )  \label{a134} \\
&&+\sum_{i=1}^{M}\int f_{i,M+1}^{(M+1)}(\mu _{A},\mu _{B};\tau )\,T(\mu
_{A},\mu _{B};\mu _{i},\mu _{C})\,d\mu _{A}\,d\mu _{B}\,d\mu _{C}  \nonumber
\end{eqnarray}
where $M=1,2,...,\infty $. This is an infinite chain of coupled differential
equations. If we invoke
\begin{equation}
f^{(M)}(\mu _{1},\mu _{2},...,\mu _{M};\tau )=f^{(1)}(\mu _{1}\mathbf{;}\tau
)\,f^{(1)}(\mu _{2}\mathbf{;}\tau )....f^{(1)}(\mu _{M}\mathbf{;}\tau ).
\label{a135}
\end{equation}
in the eq.(\ref{a134}) all the equations in the infinite chain are satisfies
provided $f^{(1)}(\mu \mathbf{;}\tau )$ satisfies eq. (\ref{a130}). This
proves that in the limit $N\rightarrow \infty $ the AMC remains valid for
all $\tau $ if we start from an uncorrelated initial state.

What happens if we start from a correlated state that does not satisfy AMC?
For finite $N$ there are always some correlations to any order. We know that
the system evolves towards microcanonical distribution. In the limit $%
N\rightarrow \infty $ microcanonical distribution obeys AMC. This means even
if we start from a correlated state the system will satisfy AMC better and
better as the system evolves towards equilibrium for large $N.$ Collisions
destroys correlations and It should take only a few collisions per particle
to destroy initial correlations. Moreover in the practical applications of
DSMC in gas dynamics the $N$ is almost always large and initial state is
chosen as almost uncorrelated from the beginning. Therefore using the first
equation to determine the single particle probability density is a
justifiable process.

\subsection{Collision invariants and the H-theorem}

We now show that expectation value $\left\langle g(\mu \mathbf{)}%
\right\rangle $ of a collision invariant $g(\mu )$ is conserved. The $g(\mu
) $ is a collision invariant if
\begin{equation}
\Delta g=g(\mu )+g(\mu _{C})-g(\mu _{A})-g(\mu _{B})=0.  \label{a140}
\end{equation}
Multiplying eq.(\ref{a130}) and integrating over $\mu $ we obtain
\begin{equation}
\frac{d}{d\tau }\int f(\mu \mathbf{)\,}g(\mu \mathbf{)\,}d\mu =\int
[f,f]\,\,g(\mu \mathbf{)}T(\mu _{A},\mu _{B};\mu _{C},\mu )\,\,d\mu
_{A}\,d\mu _{B}\,d\mu _{C}\,d\mu .  \label{a150}
\end{equation}
Using symmetries of $T(\mu _{A},\mu _{B};\mu _{C},\mu )$ and relabeling
integration variables among themselves we can write this as
\begin{equation}
\frac{d}{d\tau }\left\langle g(\mu \mathbf{)}\right\rangle =\frac{1}{4}\int
[f,f]\,\Delta g\,T(\mu _{A},\mu _{B};\mu _{C},\mu )\,d\mu _{A}\,d\mu
_{B}\,d\mu _{C}\,d\mu .  \label{a160}
\end{equation}
The integral is zero because of eq.(\ref{a140})$.$

We can derive an H-theorem for the first equation. Defining $H(\tau )$ a
\begin{equation}
H(\tau )=\int f(\mu )\,\ln (f(\mu ))\,d\mu \mathbf{,}  \label{b10}
\end{equation}
and using the eqs. (\ref{a10},\ref{a20}) and eq.(\ref{a130}) we can express $%
dH/d\tau $ as
\begin{equation}
\frac{dH}{d\tau }=-\frac{1}{4}\int \Phi [f]\,T(\mu _{A},\mu _{B};\mu
_{C},\mu )\,\,d\mu _{A}\,d\mu _{B}\,d\mu _{C}\,d\mu \mathbf{,}  \label{b20}
\end{equation}
where
\begin{equation}
\Phi [f]=\left[ f(\mu _{A})\,\,f(\mu _{B})-f(\mu )\,\,f(\mu _{C})\right]
\left[ \ln f(\mu _{A})\,\,f(\mu _{B})-\ln f(\mu )\,\,f(\mu _{C})\right] .
\label{b31}
\end{equation}
The $\Phi [f]$ can be shown to be always nonnegative as done in all kinetic
theory books and $T(\mu _{A},\mu _{B};\mu _{C},\mu )$ is intrinsically
positive. Therefore $dH/d\tau $ is nonpositive. There are two possibilities
here. The $H$ keeps decreasing toward negative infinity or it approaches an
absolute minimum asymptotically and the system approaches toward an
equilibrium distribution. Following the usual arguments of the H-theorem,
the decrease of $H$ stops only when
\begin{equation}
\ln f(\mu _{A})+\ln f(\mu _{B})=\ln f(\mu _{C})+\ln f(\mu ),  \label{b40}
\end{equation}
is satisfied which implies that $\ln f(\mu )$ is a collision invariant. If
we choose the $T(\mu _{A},\mu _{B};\mu _{C},\mu )$ such that there are
collision invariants $g_{i}(\mu )$ $(i=1,2,...,L)$ then $\ln f(\mu )$ must
be expressible as a linear combinations of these collision invariants as
\begin{equation}
\ln f(\mu )=c_{1}g_{1}(\mu )+c_{2}g_{2}(\mu )+...+c_{L}g_{L}(\mu ),
\label{b50}
\end{equation}
where $c_{1},...,c_{L}$ are parameters describing the equilibrium.

There is at least one trivial collision invariant. It is the number of
particles entering and exiting the collision which corresponds to $g_{1}(\mu
)=1$. When there are additional collision invariants the $H$ has a lower
bound usually. For the case of real gases momentum and energy are collision
invariants and this makes $H$ bounded from below.

\subsection{Example: A game of discrete money gambling}

Here we give a simple example of a direct simulation money game with finite
number of discrete states. Suppose everybody is given some random amount of
money at the beginning. Everybody in the assembly has one, two or three
dollars in their pocket. The random assignment of initial money ensures
assumption of molecular chaos from the beginning. The collisions takes place
as follows: Player 1 and player 2 share their total money such that nobody
gets more than three dollars and both players get at least one dollar. All
the possibilities satisfying these conditions have equal probabilities. If
they have total two dollars (one dollar each) then the only possibility is
that they will have one dollar each at the end with unity probability. If
they have total three dollars then the possible outcomes are (1,2) and (2,1)
with equal 1/2 probabilities. If they have total four dollars then possible
outcomes are (1,3), (3,1), (2,2) with 1/3 probability each. If they have
total five dollars then possible outcomes are (2,3) and (3,2) with 1/2
probability each. Finally if they have total six dollars (three dollars
each) then the only possibility is (3,3) with unity probability.

For this game the money is conserved in collisions and transitions between
states with equal amount of total money is possible only. For $N$ particles
the total money can have values between $N$ to $3N$ and there are a total of
$2N+1$ separate regions in phase space. One cannot cross from one to another
of these regions by making collisions.

Now that we defined the game, how does single particle distribution evolves
as we make collisions? The state variable $\mu $ is the amount of the money
in the persons pocket and it takes the values 1,2,3. Let $P_{\mu }(\tau )$
be the probability that a chosen person will have the money $\mu $ at the
collision time $\tau .$ From eq.(\ref{a131}) the $P_{\mu }(\tau )$ satisfies
\begin{eqnarray}
\frac{dP_{1}}{d\tau } &=&-P_{1}+P_{1}^{2}\,\,T(1,1,1,1)+P_{1}P_{2}\,%
\,T(1,2;2,1)  \label{c10} \\
&&+P_{2}P_{1}\,\,T(2,1;2,1)+P_{1}P_{3}\,\,T(1,3;3,1)  \nonumber \\
&&+P_{3}P_{1}\,\,T(3,1;3,1)+P_{2}P_{2}\,\,T(2,2;2,1),  \nonumber
\end{eqnarray}
\begin{eqnarray}
\frac{dP_{2}}{d\tau } &=&-P_{2}+P_{2}^{2}\,\,T(2,2,2,2)  \label{c20} \\
&&+P_{1}P_{2}\,\,T(1,2;1,2)+P_{2}P_{1}\,\,T(2,1;1,2)  \nonumber \\
&&+P_{1}P_{3}\,\,T(1,3;2,2)+P_{3}P_{1}\,\,T(3,1;2,2)  \nonumber \\
&&+P_{2}P_{3}\,\,T(2,3;3,2)+P_{3}P_{2}\,\,T(3,2;3,2),  \nonumber
\end{eqnarray}
and
\begin{eqnarray}
\frac{dP_{3}}{d\tau } &=&-P_{3}+P_{1}P_{3}\,\,T(1,3;1,3)+P_{3}P_{1}\,%
\,T(3,1;1,3)  \label{c30} \\
&&+P_{3}^{2}\,\,T(3,3;3,3)+P_{2}P_{3}\,\,T(2,3;2,3)  \nonumber \\
&&+P_{3}P_{2}\,\,T(3,2;2,3)+P_{2}^{2}\,\,T(2,2,1,3).  \nonumber
\end{eqnarray}
Inserting the $T$ values this can be written as
\begin{eqnarray}
\frac{dP_{1}}{d\tau } &=&-P_{1}+P_{1}^{2}+P_{1}P_{2}+\frac{2}{3}P_{1}P_{3}+%
\frac{1}{3}P_{2}^{2},  \label{c40} \\
\frac{dP_{2}}{d\tau } &=&-P_{2}+\frac{1}{3}P_{2}^{2}+P_{1}P_{2}+\frac{2}{3}%
P_{1}P_{3}+P_{2}P_{3},  \label{c41} \\
\frac{dP_{3}}{d\tau } &=&-P_{3}+\frac{2}{3}P_{1}P_{3}+P_{2}P_{3}+P_{3}^{2}+%
\frac{1}{3}P_{2}^{2}.  \label{c42}
\end{eqnarray}

This is a complicated set of nonlinear differential equations. But there are
simplifying features because we know the collision invariants $g_{1}(\mu )=1$
and $g_{2}(\mu )=\mu $. Summing the eqs.(\ref{c40},\ref{c41},\ref{c42}) we
obtain
\begin{equation}
\frac{d}{d\tau }\left( P_{1}+P_{2}+P_{3}\right) =\left(
P_{1}+P_{2}+P_{3}-1\right) \left( P_{1}+P_{2}+P_{3}\right) ,  \label{c46}
\end{equation}
and
\begin{equation}
\frac{d}{d\tau }\left( P_{1}+2P_{2}+3P_{3}\right) =\left(
P_{1}+P_{2}+P_{3}-1\right) \left( P_{1}+2P_{2}+3P_{3}\right) .  \label{c46b}
\end{equation}
The first equation tells us that since $P_{1}+P_{2}+P_{3}=1$ at the
beginning it always remains unity and probability is conserved. The second
equation tells us that since $P_{1}+P_{2}+P_{3}-1=0$ always the expectation
value $\left\langle \mu \right\rangle =P_{1}+2P_{2}+3P_{3}$ is conserved.

We denote expected money in the pocket with $m$. We have two equations
\begin{eqnarray}
P_{1}+P_{2}+P_{3} &=&1,  \label{c47} \\
P_{1}+2P_{2}+3P_{3} &=&m,  \label{c48}
\end{eqnarray}
from which we solve $P_{2}$ and $P_{3}$ as
\begin{eqnarray}
P_{2} &=&-2P_{1}+3-m,  \label{c49} \\
P_{3} &=&P_{1}+m-2.  \label{c50}
\end{eqnarray}
Inserting $P_{2}$ and $P_{3}$ in the eq.(\ref{c40}) we obtain
\begin{equation}
\frac{dP_{1}}{d\tau }=P_{1}^{2}+(m-\frac{10}{3})P_{1}+\frac{1}{3}(3-m)^{2}.
\label{ek10}
\end{equation}
Calculating roots of the quadratic term on the right we write this as
\begin{equation}
\frac{dP_{1}}{d\tau }=\left( P_{1}-r_{1}\right) \left( P_{1}-r_{2}\right) ,
\label{ek20}
\end{equation}
where $r_{1}$ and $r_{2}$ are
\begin{eqnarray}
r_{1} &=&\frac{1}{6}\left( 10-3m+\sqrt{1+3(m-1)(3-m)}\right) ,  \label{ek30}
\\
r_{2} &=&\frac{1}{6}\left( 10-3m-\sqrt{1+3(m-1)(3-m)}\right) .  \label{ek40}
\end{eqnarray}
Notice that since $1\leq m\leq 3$ the term under the square root is always
greater than or equal to unity.

Solving eq.(\ref{ek20}) is straightforward and we obtain
\begin{equation}
P_{1}(\tau )=\frac{r_{2}(p_{0}-r_{1})-r_{1}(p_{0}-r_{2})\mathrm{e}^{-\lambda
\tau }}{(p_{0}-r_{1})-(p_{0}-r_{2})\mathrm{e}^{-\lambda \tau }},
\label{ek50}
\end{equation}
where $p_{0}=P_{1}(\tau =0)$ and $\lambda =r_{1}-r_{2}$. It is easy to
verify that $P_{1}(\infty )=r_{2}$ and $P_{1}(\tau )\,$approaches this limit
exponentially fast. One can check from eq.(\ref{ek40}) that $r_{2}=1$ at $%
m=1 $ and $r_{2}=0$ at $m=3$ and it behaves as it is expected.

The conditions $0\leq P_{2}\leq 1$ and $0\leq P_{3}\leq 1$ together with
eqs.(\ref{c49},\ref{c50}) gives conditions that $P_{1}(\tau )$ must satisfy.
These conditions are expressed as $2-m\leq P_{1}\leq (3-m)/2$ when $m\leq 2$
and $0\leq P_{1}\leq (3-m)/2$ when $m>2$. Therefore $P_{1}(\tau =0)\,$%
initial value should obey these limitations.

To find the equilibrium distribution directly without solving the
differential equation we set $dP_{\mu }/d\tau =0$ for $\mu =1,2,3$ in eqs.(%
\ref{c40},\ref{c41},\ref{c42}) and we obtain a set of algebraic nonlinear
equations. Setting $P_{1}=a$, $P_{2}=ab$, $P_{3}=ab^{2}$ all three equations
are satisfied provided the normalization condition
\begin{equation}
a(1+b+b^{2})=1,  \label{c59}
\end{equation}
holds. We were able to guess this solution from the H-theorem. There are two
collision invariants $g_{1}=1$ and $g_{2}(\mu )=\mu $. The second one is a
result of conservation of money in the collisions. Therefore according to
the H-theorem we must have $\ln P_{\mu }=C_{1}+C_{2}\mu $ and this gives the
solution $P_{\mu }=ab^{\mu -1}$. We need one more relation to determine both
$a$ and $b$. This comes from expected money in the pocket:
\begin{equation}
m=a\left( 1+2b+3b^{2}\right) ,  \label{c60}
\end{equation}
which is a conserved quantity during the 'time' evolution and it is set by
the initial conditions. Solving these two equation we obtain
\begin{eqnarray}
a &=&\frac{1}{6}\left( 10-3m-\sqrt{1+3(m-1)(3-m)}\right) ,  \label{c65} \\
b &=&\left( m-2+\sqrt{1+3(m-1)(3-m)}\right) /2(3-m).  \nonumber
\end{eqnarray}
Notice that $a=r_{2}$ and this agrees with solution of the differential
equation.

The H-function
\begin{equation}
H=P_{1}\ln P_{1}+P_{2}\ln P_{2}+P_{3}\ln P_{3},  \label{c70}
\end{equation}
is bounded from below for this problem since the function $x\ln x$ is
bounded from below and $0\leq P_{\mu }\leq 1$. We minimize $H$ with the
constraint that the expected money is fixed and probabilities are
normalized. The constraints can be adopted with Lagrange multipliers. Taking
the auxiliary function
\begin{eqnarray}
\Psi &=&P_{1}\ln P_{1}+P_{2}\ln P_{2}+P_{3}\ln P_{3}  \label{c80} \\
&&-\lambda _{2}(P_{1}+P_{2}+P_{3}-1)-\lambda _{2}(P_{1}+2P_{2}+3P_{3}-m),
\nonumber
\end{eqnarray}
and setting $\partial \Psi /\partial P_{1}=\partial \Psi /\partial
P_{2}=\partial \Psi /\partial P_{3}=0$ we obtain the same solution $P_{\mu
}=ab^{\mu -1}$ where $a$ and $b$ satisfies the eqs.(\ref{c59},\ref{c60}).
The minimum value of H becomes
\begin{equation}
H=a\ln a+ab\ln ab+ab^{2}\ln ab^{2}=\ln (ab^{m-1}).  \label{c90}
\end{equation}

\subsection{Example2: A game of continuous money gambling}

Here we give another example of direct simulation money games with
continuous states. In this case we were not even able to solve one particle
probability distribution. We just find the equation for one particle
distribution and guess the stationary one particle distribution from the
H-theorem. We then show that it satisfies the equation for single particle
probability equation.

This time initially we give players a random amount of money between zero
and, say, ten dollars. Suppose we pick a pair to collide. player1 has $\mu
_{1}$ and player2 has $\mu _{2}$ amount of money. A computer produces a
random number $p$ between zero and one. Player1 takes $p(\mu _{1}+\mu _{2})$
and player2 takes $(1-p)(\mu _{1}+\mu _{2})$ amounts of money and we pick
another pair to collide. What is the final distribution when the system
comes to equilibrium?

The probability distribution that a person will have money $\mu $ satisfies
the eq.(\ref{a131})
\begin{equation}
\frac{\partial f(\mu \mathbf{)}}{\partial \tau }=-f(\mu )+\int_{0}^{\infty
}da\int_{0}^{\infty }db\,f(a)\,f(b)\,T(a,b,\mu ,\nu )\,da\,db\,d\nu ,
\label{d10}
\end{equation}
where the collision kernel is
\begin{equation}
T(a,b,\mu ,\nu )=\frac{1}{a+b}\delta (a+b-\mu -\nu )\Theta (a)\,\Theta
(b)\,\Theta (\mu )\,\Theta (\nu ).  \label{d21}
\end{equation}
Here $\Theta (x)$ is the standard step function
\begin{equation}
\Theta (x)=\left\{
\begin{array}{ll}
0\quad & {}x<0 \\
1\quad & x\geq 0
\end{array}
\right. .  \label{d30}
\end{equation}
If we insert the $T(a,b,\mu ,\nu )$ given in the eq.(\ref{d21}) into the eq.(%
\ref{d10}) and perform the $\nu $ integral we obtain
\begin{equation}
\frac{\partial f(\mu \mathbf{)}}{\partial \tau }=-f(\mu )+\int_{0}^{\infty
}da\int_{0}^{\infty }db\,\Theta (a+b-\mu )\,\frac{f(b)\,f(a)}{a+b}.
\label{d40}
\end{equation}
This can be further simplified by changing variables $x=a+b$ and $y=a$ which
yields
\begin{equation}
\frac{\partial f(\mu \mathbf{)}}{\partial \tau }=-f(\mu )+\int_{\mu
}^{\infty }dx\int_{0}^{x}dy\,\frac{f(y)\,f(x-y)}{x}.  \label{d50}
\end{equation}

The H-theorem insures that this equation will converge to an equilibrium
distribution as $\tau \rightarrow \infty $. Since we have money conservation
in the collisions there are two collision invariants $g_{1}(\mu )=1$ and $%
g_{2}(\mu )=\mu $. Then the equilibrium distribution is
\begin{equation}
f_{eq}(\mu )=A\,e^{-B\mu }.  \label{d60}
\end{equation}
If the average money initially given to each person is $m$, the $f(\mu )$
should satisfy two conditions
\begin{eqnarray}
\int_{0}^{\infty }f(\mu )\,d\mu &=&1,  \label{d80} \\
\int_{0}^{\infty }\mu \,f(\mu )\,d\mu &=&m,  \label{d81}
\end{eqnarray}
and they fix the values of $A$ and $B$ in the eq.(\ref{d60}). The solution
is
\begin{equation}
f_{eq}(\mu )=\frac{1}{m}\,e^{-\mu /m}.  \label{d90}
\end{equation}
If we insert this solution into eq.(\ref{d50}) we can easily check that
right side of the equation becomes zero which confirms that $f_{eq}(\mu )$
is the equilibrium distribution.

\section{ Application of the direct simulation formalism to homogenous gases}

\subsection{Center of mass frame}

In the following sections we will need some results from studying the
collision in the center of mass frame. Instead of deriving them for each
case separately we derive the relevant results once for the most general
case in this subsection and refer to formulae derived here as needed in the
following subsections. In the rest of the paper bold letters denote vector
quantities.

Particles with states $\mu _{A}=\mathbf{v}_{A}\mathbf{\ }$and $\mu _{B}=%
\mathbf{v}_{B}$ and enter the collision and particles with states $\mu _{C}=%
\mathbf{v}_{C}$ and $\mu _{D}=\mathbf{v}$ exit the collision. We define the
center of mass (CM) coordinates as
\begin{eqnarray}
\mathbf{H} &=&(m_{A}\mathbf{v}_{A}+m_{B}\mathbf{v}_{B})/(m_{A}+m_{B})
\label{e3} \\
\mathbf{H}^{\prime } &=&(m_{A}\mathbf{v}_{C}+m_{B}\mathbf{v})/(m_{A}+m_{B}),
\label{e3b}
\end{eqnarray}
and
\begin{equation}
\begin{array}{lll}
\mathbf{u=v}_{A}-\mathbf{v}_{B},\;\;\; & u=\left| \mathbf{u}\right| , &
\mathbf{n}=\mathbf{u}/u \\
\mathbf{u}^{\prime }\mathbf{=v}_{C}-\mathbf{v}, & u^{\prime }=\left| \mathbf{%
u}^{\prime }\right| ,\;\;\; & \mathbf{n}^{\prime }=\mathbf{u}^{\prime
}/u^{\prime }
\end{array}
\label{e4}
\end{equation}
where $m_{A}$ is the mass of particles $A$ and $C$ and $m_{B}$ is the mass
of particles $B$ and $D$. For one kind of gas all masses are equal and
formulae for CM velocities $\mathbf{H}$ and $\mathbf{H}^{\prime }$ reduce to
\begin{equation}
\begin{array}{ll}
\mathbf{H}=(\mathbf{v}_{A}+\mathbf{v}_{B})/2,\;\;\; & \mathbf{H}^{\prime }=(%
\mathbf{v}_{C}+\mathbf{v})/2.
\end{array}
\label{e5}
\end{equation}
Integrations over $\mathbf{v}_{A}$ and $\mathbf{v}_{B}$ can be carried over
in the variables $\mathbf{H}$ and $\mathbf{u}$. The transformation between
these two sets of variables are linear and the Jacobian is unity. Therefore
\begin{equation}
\int f(\mathbf{v}_{A},\mathbf{v}_{B})d^{3}\mathbf{v}_{A}d^{3}\mathbf{v}%
_{B}=\int f(\mathbf{H},\mathbf{u})d^{3}\mathbf{H}d^{3}\mathbf{u.}
\label{e34}
\end{equation}
In the following subsections we will deal with integrations over $\mathbf{v}%
_{A}$, $\mathbf{v}_{B}$, $\mathbf{v}_{C}$. Integrations over $\mathbf{v}_{A}$%
, $\mathbf{v}_{B}$ will be converted to integration over $\mathbf{H}$ and $%
\mathbf{u}$ in the CM\ frame. In each case there will be a Dirac delta
function removing the integral over $\mathbf{H}$. Integration over $\mathbf{v%
}_{C}$ will be converted to integration over $\mathbf{u}^{\prime }$ since $%
\mathbf{v}_{C}=\mathbf{u}^{\prime }\mathbf{+v}$ and there is no integration
over $\mathbf{v.}$ Furthermore integrations over $\mathbf{u}^{\prime }$ will
be carried in spherical coordinates as
\begin{equation}
\int f(\mathbf{u}^{\prime })d^{3}\mathbf{u}^{\prime }=\int f(\mathbf{u}%
^{\prime })(u^{\prime })^{2}du^{\prime }d\mathbf{n}^{\prime }  \label{e36}
\end{equation}
and in each case there will be a Dirac delta function removing the
integration over $u^{\prime }.$ In the final expressions the integration
over solid angle $\mathbf{n}^{\prime }$ and $\mathbf{u}$ remain at the end.

In order to evaluate the integrals we will encounter in the following
subsections we must express $\mathbf{v}_{A},\mathbf{v}_{B},\mathbf{v}_{C}$
in terms of the variables $\mathbf{v,u,n}^{\prime }.$ This is a simple
exercise in collision kinetics. We will do this for the inelastic collisions
with unequal masses. This is the most general case we will deal in this
paper. We will assume that molecules have internal energies $\epsilon (A),$ $%
\epsilon (B)$ and $\epsilon (C),$ $\epsilon (D)$. Let $\epsilon =\epsilon
(A)+\epsilon (B)$ and $\epsilon ^{\prime }=\epsilon (C)+\epsilon (D)$. From
energy conservation we have $u^{\prime }(u)=\sqrt{u^{2}+2(\epsilon -\epsilon
^{\prime })/m_{r}}$ where $m_{r}=m_{A}m_{B}/(m_{A}+m_{B})$ is the reduced
mass and $m_{A}$, $m_{B}$ are masses of the colliding particles. We can
write $\mathbf{u}^{\prime }=u^{\prime }(u)\mathbf{n}^{\prime }$ and $\mathbf{%
v}_{C}=\mathbf{v+}u^{\prime }(u)\mathbf{n}^{\prime }$. From CM velocity
conservation we have
\begin{equation}
m_{A}\mathbf{v}_{A}+m_{B}\mathbf{v}_{B}=m_{A}\mathbf{v}_{C}+m_{B}\mathbf{v=}%
(m_{A}+m_{B}\mathbf{)v}+m_{A}u^{\prime }(u)\mathbf{n}^{\prime }  \label{e93}
\end{equation}
and we also have $\mathbf{v}_{A}-\mathbf{v}_{B}=\mathbf{u}.$ We solve $%
\mathbf{v}_{A}$, $\mathbf{v}_{B}$, $\mathbf{v}_{C}$ from these as
\begin{eqnarray}
\mathbf{v}_{A} &=&\mathbf{v}+\frac{m_{A}}{m_{A}+m_{B}}u^{\prime }(u)\mathbf{n%
}^{\prime }+\frac{m_{B}}{m_{A}+m_{B}}\mathbf{u}  \label{e100} \\
\mathbf{v}_{B} &=&\mathbf{v}+\frac{m_{A}}{m_{A}+m_{B}}u^{\prime }(u)\mathbf{n%
}^{\prime }-\frac{m_{A}}{m_{A}+m_{B}}\mathbf{u}  \label{e101} \\
\mathbf{v}_{C} &=&\mathbf{v+}u^{\prime }(u)\mathbf{n}^{\prime }  \label{e102}
\\
u^{\prime }(u) &=&\sqrt{u^{2}+2(\epsilon -\epsilon ^{\prime })/m_{r}}
\label{e103}
\end{eqnarray}

For one kind of gas ($m_{A}=m_{B}=m$ ) without internal states ($\epsilon
(A)=\epsilon (B)=\epsilon (C)=\epsilon (D)=0$) these equations reduce to
\begin{eqnarray}
\mathbf{v}_{A} &=&\mathbf{v}+(u\mathbf{n}^{\prime }+\mathbf{u)/2}
\label{e105} \\
\mathbf{v}_{B} &=&\mathbf{v}+(u\mathbf{n}^{\prime }-\mathbf{u)/2}
\label{e106} \\
\mathbf{v}_{C} &=&\mathbf{v+}u\mathbf{n}^{\prime }  \label{e107}
\end{eqnarray}
Again for one kind of gas ($m_{A}=m_{B}=m$ and $m_{r}=m/2$) with internal
states eqs.(\ref{e100},\ref{e101},\ref{e102},\ref{e103}) reduce to
\begin{eqnarray}
\mathbf{v}_{A} &=&\mathbf{v}+\left[ u^{\prime }(u)\mathbf{n}^{\prime }+%
\mathbf{u}\right] \mathbf{/}2  \label{g75} \\
\mathbf{v}_{B} &=&\mathbf{v}+\left[ u^{\prime }(u)\mathbf{n}^{\prime }-%
\mathbf{u}\right] /2  \label{g76} \\
\mathbf{v}_{C} &=&\mathbf{v+}u^{\prime }(u)\mathbf{n}^{\prime }  \label{g77}
\\
u^{\prime }(u) &=&\sqrt{u^{2}+4(\epsilon -\epsilon ^{\prime })/m}
\label{g78}
\end{eqnarray}
For a mixture of gases without internal states eqs.(\ref{e100},\ref{e101},%
\ref{e102},\ref{e103}) reduce to
\begin{eqnarray}
\mathbf{v}_{A} &=&\mathbf{v}+\frac{m_{A}}{m_{A}+m_{B}}\,u\mathbf{n}^{\prime
}+\frac{m_{B}}{m_{A}+m_{B}}\,\mathbf{u,}  \label{f80} \\
\mathbf{v}_{B} &=&\mathbf{v}+\frac{m_{A}}{m_{A}+m_{B}}\,u\mathbf{n}^{\prime
}-\frac{m_{A}}{m_{A}+m_{B}}\,\mathbf{u,}  \label{f81} \\
\mathbf{v}_{C} &=&\mathbf{v+}u\mathbf{n}^{\prime }.  \label{f82}
\end{eqnarray}
And for a mixture of gases with internal states eqs.(\ref{e100},\ref{e101},%
\ref{e102},\ref{e103}) are the formule.

\subsection{One kind of gas without internal degrees of freedom}

The state of particles are defined by three components of the velocity
vector $\mathbf{v}$. (We use bold letters for vectors throughout this paper)
Bird's original algorithm to keep track of time in the simulation was the
'time counter method'. Later Bird introduced 'No time counter method' (NTC)
and declared time counter method 'obsolete' in his book.\cite{Bird94} Time
counter method is more difficult (if not impossible) to formulate in the
direct simulation formalism given in this paper and since NTC is the
algorithm currently used we will derive NTC algorithms only in this paper.

Here the state index $\mu $ refer the velocity vectors and the integration
over $\mu $ stands for three integrations over components of velocities. The
NTC\ kernel $S(\mathbf{v}_{A},\mathbf{v}_{B};\mathbf{v}_{C},\mathbf{v}%
)=S_{1}+S_{2}$ is given by
\begin{eqnarray}
S_{1} &=&\frac{2}{R}\delta \left( \mathbf{H}-\mathbf{H}^{\prime }\right)
\,\delta \left( u^{2}-(u^{\prime })^{2}\right) \,\sigma (\mathbf{n},\mathbf{n%
}^{\prime })  \label{e20} \\
S_{2} &=&\left( 1-\frac{u\Sigma }{R}\right) \,\delta \left( \mathbf{v}_{C}-%
\mathbf{v}_{A}\right) \,\delta \left( \mathbf{v}-\mathbf{v}_{B}\right)
\label{e30}
\end{eqnarray}
Here $\sigma (\mathbf{n},\mathbf{n}^{\prime })$ is the differential cross
section and $\Sigma \,$is the total cross section which is given by
\begin{equation}
\Sigma =\int \sigma (\mathbf{n},\mathbf{n}^{\prime })\,d\mathbf{n}^{\prime },
\label{e40}
\end{equation}
where $d\mathbf{n}^{\prime }$ is the solid angle in the direction of $%
\mathbf{n}^{\prime }$. The $\sigma (\mathbf{n},\mathbf{n}^{\prime })$
depends on the angle $\theta $ between $\mathbf{n}$ and $\mathbf{n}^{\prime
} $ ($\mathbf{n}^{\prime }\cdot \mathbf{n}=\cos \theta $). Hence $\sigma (%
\mathbf{n},\mathbf{n}^{\prime })=\sigma (\mathbf{n}^{\prime },\mathbf{n})$
and the kernel is obviously symmetric. The term $\delta (u^{2}-(u^{\prime
})^{2})=\delta (u-u^{\prime })/2u$ represents energy conservation and $%
\delta \left( \mathbf{H}-\mathbf{H}^{\prime }\right) $ represents
conservation of center of mass (CM) velocity which is the same thing as the
conservation of momentum. The kernel satisfies the normalization condition
\begin{equation}
\int S(\mathbf{v}_{A},\mathbf{v}_{B};\mathbf{v}_{C},\mathbf{v})\,d^{3}%
\mathbf{v}_{C}\,d^{3}\mathbf{v=\int }S(\mathbf{v}_{A},\mathbf{v}_{B};\mathbf{%
v}_{C},\mathbf{v})\,d^{3}\mathbf{H}^{\prime }\,d^{3}\mathbf{u}^{\prime }=1.
\label{e50}
\end{equation}
Here the integral is taken in the CM coordinates. The Jacobian of the CM
transformation is unity and $d^{3}\mathbf{u}^{\prime }=(u^{\prime
})^{2}du^{\prime }d\mathbf{n}^{\prime }$.

The $S_{2}$ part of the kernel directly transfer initial velocities to the
final velocities with a probability $\left( 1-u\Sigma /R\right) $ and hence
causes a null collision. A null collision is a collision that particles do
not change their states. The probability of making a real collision is
\begin{equation}
\int S_{1}(\mathbf{v}_{A},\mathbf{v}_{B};\mathbf{v}_{C},\mathbf{v})\,d^{3}%
\mathbf{v}_{C}\,d^{3}\mathbf{v=}\frac{u\Sigma }{R}  \label{e60}
\end{equation}
where integral is calculated in the CM coordinates.

Inserting $S(\mathbf{v}_{A},\mathbf{v}_{B};\mathbf{v}_{C},\mathbf{v})$ in
eq.(\ref{a130}) we obtain
\begin{equation}
\frac{\partial f(\mathbf{v})}{\partial \tau }=\int [f,f]\,\,S_{1}(\mathbf{v}%
_{A},\mathbf{v}_{B};\mathbf{v}_{C},\mathbf{v})\,d^{3}\mathbf{v}_{A}\,d^{3}%
\mathbf{v}_{B}\,d^{3}\mathbf{v}_{C}.  \label{e70}
\end{equation}
where
\begin{equation}
\lbrack f,f]=f(\mathbf{v}_{A})\,f(\mathbf{v}_{B})-f(\mathbf{v}_{C})\,f(%
\mathbf{v}),  \label{e72}
\end{equation}
The $S_{2}\,$part of the kernel gives zero contribution in the collision
integral
\begin{equation}
\int [f,f]\,\,\delta \left( \mathbf{v}_{C}-\mathbf{v}_{A}\right) \,\delta
\left( \mathbf{v}-\mathbf{v}_{B}\right) \,d^{3}\mathbf{v}_{A}\,d^{3}\mathbf{v%
}_{B}\,d^{3}\mathbf{v}_{C}=0.  \label{e80}
\end{equation}

We evaluate the integral in eq.(\ref{e70}) in the CM coordinates. We write $%
d^{3}\mathbf{v}_{A}\,d^{3}\mathbf{v}_{B}=d^{3}\mathbf{H}\,d^{3}\mathbf{u}$
and $d^{3}\mathbf{v}_{C}=d^{3}\mathbf{u}^{\prime }=(u^{\prime
})^{2}\,du^{\prime }\,d\mathbf{n}^{\prime }$. When we do the integral we
obtain
\begin{equation}
\frac{\partial f(\mathbf{v})}{\partial \tau }=\frac{1}{R}\int
[f,f]\,u\,\sigma (\mathbf{n},\mathbf{n}^{\prime })\,d^{3}\mathbf{u}\,d%
\mathbf{n}^{\prime },  \label{e90}
\end{equation}
where $\mathbf{v}_{A},\mathbf{v}_{B},\mathbf{v}_{C}$ are expressed in terms
of the variables $\mathbf{v,u,n}^{\prime }$ in eqs.(\ref{e105},\ref{e106},%
\ref{e107}).

The equation(\ref{e90}) is essentially the Boltzmann equation with the
difference that the Boltzmann equation is written for density in physical
space. To obtain the Boltzmann equation we write this equation for $F(%
\mathbf{v})=\left( N/V\right) f(\mathbf{v})$ where $V\,$is the volume of the
gas. Then we obtain
\begin{equation}
\frac{\partial F(\mathbf{v})}{\partial \tau }=\frac{1}{R}\left( \frac{V}{N}%
\right) \int \left[ F(\mathbf{v}_{A})F(\mathbf{v}_{B})-F(\mathbf{v}_{C})F(%
\mathbf{v})\right] \,u\,\sigma (\mathbf{n},\mathbf{n}^{\prime })\,d^{3}%
\mathbf{u}\,d\mathbf{n}^{\prime }  \label{e110}
\end{equation}
Now, if we change to the variable $t=\tau V/RN=2nV/RN^{2}$ we obtain the
Boltzmann equation for a homogenous gas
\begin{equation}
\frac{\partial F(\mathbf{v})}{\partial t}=\int \left[ F(\mathbf{v}_{A})\,F(%
\mathbf{v}_{B})-F(\mathbf{v}_{C})\,F(\mathbf{v})\right] \,u\,\sigma (\mathbf{%
n},\mathbf{n}^{\prime })\,d^{3}\mathbf{u}\,d\mathbf{n}^{\prime }
\label{e120}
\end{equation}
Here $t$ must be interpreted as the physical time and $t=2nV/RN^{2}$ formula
connects the physical time $t$ and number of collision attempts $n$.

Let us state the algorithm for a homogenous gas. We choose a number $R$ big
enough such that for only very few (say less than one in thousand) pairs $%
u\Sigma /R$ will exceed unity. We make $n=RN^{2}t/2V$ collision attempts to
reach the desired time. For each pair we take a random number $r$ and we
allow the collision to happen if $r<u\Sigma /R$. If the collision is
allowed, we choose the direction of scattering $\mathbf{n}^{\prime }$
according to the probability density $\sigma (\mathbf{n},\mathbf{n}^{\prime
})/\Sigma $ and a few more random numbers are used for that. Then we
calculate and store final velocities for the colliding pairs and pick
another pair. We keep taking and colliding pairs until we reach the desired
time.

Suppose the formula $n=RN^{2}t/2V$ yields 234.783 collisions. How do you
make 0.783 collisions? The way to do this in practise is to make 234
collisions first. Then throw a random number $r$ and if $r<0.783$ then go on
to make a collision attempt. This can be justified from the formula
\begin{equation}
f(\mu \mathbf{;}n+1)=f(\mu \mathbf{;}n)+\frac{2}{N}\int [f,f]\,T(\mu
_{A},\mu _{B};\mu _{C},\mu )\,d\mu _{A}\,d\mu _{B}\,d\mu _{C}.  \label{e131}
\end{equation}
After making $n$ collision attempts with the NTC\ kernel $S(\mathbf{v}_{A},%
\mathbf{v}_{B};\mathbf{v}_{C},\mathbf{v})\,$we can change the kernel to
\begin{equation}
P(\mathbf{v}_{A},\mathbf{v}_{B};\mathbf{v}_{C},\mathbf{v})=q\,S(\mathbf{v}%
_{A},\mathbf{v}_{B};\mathbf{v}_{C},\mathbf{v})+(1-q)\,\delta \left( \mathbf{v%
}_{C}-\mathbf{v}_{A}\right) \,\delta \left( \mathbf{v}-\mathbf{v}_{B}\right)
.  \label{e140}
\end{equation}
This kernel makes a NTC\ collision attempt with a probability $q\,$(which
was $0.783$ in the above example) and a null collision happens with the
probability $1-q$. We use this kernel for the $(n+1)^{th}$ collision attempt
(it is permissible to change the kernel) and this causes another $\Delta
\tau =2q/N\,$collision time and $\Delta t=q(2V/RN^{2})$ real time increase.

\subsection{Mixture of gases without internal degrees of freedom}

The state of particles are defined by three components of the velocity
vector $\mathbf{v}$ and one kind index for which we will use $p,q,r,s$
characters. We have $M$ kind of gas without internal states in the mixture
and there are $N_{p}$ number of molecules of the $p^{th}$ kind. The mass of $%
p^{th}$ kind molecule is $m_{p}$. The probability density $f(\mu )=f(\mathbf{%
v,}p)$ will be written as $f^{p}(\mathbf{v})$.

Particles with states $\mu _{A}=(\mathbf{v}_{A},s),\mathbf{\ }$and $\mu
_{B}=(\mathbf{v}_{B},r)$ enter the collision and particles with states $\mu
_{C}=(\mathbf{v}_{C},q)$ and $\mu _{D}=(\mathbf{v,}p\mathbf{)}$ exits the
collision. The integration over $\mu $ such as $\int f^{p}(\mathbf{v})d\mu $
stands for three integrations over $\mathbf{v}$ and summation over $p$. The
center of mass (CM) coordinates are defined in eqs.(\ref{e3},\ref{e3b},\ref
{e4}).

The NTC\ kernel $G_{pq}^{rs}(\mathbf{v}_{A},\mathbf{v}_{B};\mathbf{v}_{C},%
\mathbf{v})=G_{1}+G_{2}$ is given by
\begin{eqnarray}
G_{1} &=&\frac{2}{R}\,\delta \left( \mathbf{H}-\mathbf{H}^{\prime }\right)
\,\,\delta \left( u^{2}-(u^{\prime })^{2}\right) \,\sigma _{pq}(\mathbf{n},%
\mathbf{n}^{\prime })\,\delta _{pr}\,\delta _{qs},  \label{f20} \\
G_{2} &=&\left( 1-\frac{u\Sigma _{pq}}{R}\right) \,\delta \left( \mathbf{v}%
_{C}-\mathbf{v}_{A}\right) \,\delta \left( \mathbf{v}-\mathbf{v}_{B}\right)
\,\delta _{pr}\,\delta _{qs}.  \label{f30}
\end{eqnarray}
Here $\sigma _{pq}(\mathbf{n},\mathbf{n}^{\prime })$ is the differential
cross section between gases of the $p^{th}$ and $q^{th}$ kind and $\Sigma
_{pq}\,$is the total cross section which is given by
\begin{equation}
\Sigma _{pq}=\int \sigma _{pq}(\mathbf{n},\mathbf{n}^{\prime })\,d\mathbf{n}%
^{\prime },  \label{f40}
\end{equation}
where $d\mathbf{n}^{\prime }$ is the solid angle in the direction of $%
\mathbf{n}^{\prime }$. The $\delta _{pr}\delta _{qs}$ term in the kernel
insures that particles do not loose their identities during the collisions.
Again $\sigma _{pq}(\mathbf{n},\mathbf{n}^{\prime })=\sigma _{rs}(\mathbf{n,n%
}^{\prime })\,$due to the $\delta _{pr}\delta _{qs}$ term and we also have
the symmetry $\sigma _{pq}(\mathbf{n},\mathbf{n}^{\prime })=\sigma _{qp}(%
\mathbf{n}^{\prime },\mathbf{n})$. The kernel is obviously symmetric. The
term $\delta (u^{2}-(u^{\prime })^{2})\ $ and $\delta \left( \mathbf{H}-%
\mathbf{H}^{\prime }\right) $ have the same meanings as before and the
kernel satisfies the normalization condition
\begin{equation}
\sum_{p=1}^{M}\sum_{q=1}^{M}\int G_{pq}^{rs}(\mathbf{v}_{A},\mathbf{v}_{B};%
\mathbf{v}_{C},\mathbf{v})\,d^{3}\mathbf{v}_{C}\,d^{3}\mathbf{v}=1.
\label{f50}
\end{equation}

Again $G_{2}$ part of the kernel directly transfer initial velocities to the
final velocities with a probability $1-(u\Sigma _{rs})/R$ and hence causes a
null collision. The probability of making a real collision is
\begin{equation}
\sum_{p=1}^{M}\sum_{q=1}^{M}\int (G_{1})_{pq}^{rs}(\mathbf{v}_{A},\mathbf{v}%
_{B};\mathbf{v}_{C},\mathbf{v})\,d^{3}\mathbf{v}_{C}\,d^{3}\mathbf{v=}\frac{%
u\Sigma _{rs}}{R},  \label{f60}
\end{equation}
where integral is calculated in the CM coordinates.

Inserting $G_{pq}^{rs}(\mathbf{v}_{A},\mathbf{v}_{B};\mathbf{v}_{C},\mathbf{v%
})$ in eq.(\ref{a130}) and doing the summations over $r,s$ and doing the
integrals in the CM coordinates we obtain
\begin{eqnarray}
\frac{\partial f^{p}(\mathbf{v})}{\partial \tau } &=&\sum_{q=1}^{M}\int
G_{pq}^{pq}(\mu _{A},\mu _{B};\mu _{C},\mu )\,[f^{q},f^{p}]\,\,d^{3}\mathbf{v%
}_{A}\,d^{3}\mathbf{v}_{B}\,d^{3}\mathbf{v}_{C},  \label{f70} \\
&=&\frac{1}{R}\sum_{q=1}^{M}\int [f^{q},f^{p}]\,\,u\,\sigma _{pq}(\mathbf{n},%
\mathbf{n}^{\prime })\,d^{3}\mathbf{u\,}d\mathbf{n}^{\prime },  \label{f71}
\end{eqnarray}
where
\begin{equation}
\lbrack f^{q},f^{p}]=f^{q}(\mathbf{v}_{A})\,f^{p}(\mathbf{v}_{B})-f^{q}(%
\mathbf{v}_{C})\,f^{p}(\mathbf{v})  \label{f74}
\end{equation}

Again we write this equation for $F^{p}(\mathbf{v})=\left( N/V\right) f^{p}(%
\mathbf{v})$ and take $t=2nV/RN^{2}$ to obtain Boltzmann equation for a
mixture of homogenous gases without internal states
\begin{equation}
\frac{\partial F^{p}(\mathbf{v})}{\partial t}=\sum_{q=1}^{M}\int \left[
F^{q}(\mathbf{v}_{A})\,F^{p}(\mathbf{v}_{B})-F^{q}(\mathbf{v}_{C})\,F^{p}(%
\mathbf{v})\right] \,u\,\sigma _{pq}(\mathbf{n},\mathbf{n}^{\prime })\,d^{3}%
\mathbf{u\,}d\mathbf{n}^{\prime }.  \label{f90}
\end{equation}
Here $\mathbf{v}_{A},\mathbf{v}_{B},\mathbf{v}_{C}$ are expressed in terms
of the variables $\mathbf{v,u,n}^{\prime }$ in eqs.(\ref{f80},\ref{f81},\ref
{f82}).

The algorithm is the same. We take $n=RN^{2}t/2V$ pairs and allow each
collision with a probability $(u\Sigma _{rs})/R.$ If the collision is
allowed we choose the scattering angle according to the $\sigma _{rs}(%
\mathbf{n},\mathbf{n}^{\prime })/\Sigma _{rs}$ probability distribution.

Note that the normalization of $f^{p}(\mathbf{v})\,$is given by
\begin{equation}
\sum_{p=1}^{M}\int f^{p}(\mathbf{v})\,d^{3}\mathbf{v}=1.  \label{f100}
\end{equation}
The integral $\int f^{p}(\mathbf{v})d^{3}\mathbf{v}$ is conserved during the
simulation. From eq.(\ref{f70}) its rate of change is
\begin{eqnarray}
\frac{d}{d\tau }\int f^{p}(\mathbf{v})\,d^{3}\mathbf{v} &=&\int \frac{%
\partial f^{p}(\mathbf{v})}{\partial \tau }\,d^{3}\mathbf{v=}%
\sum_{q=1}^{M}\int G_{pq}^{pq}(\mathbf{v}_{A},\mathbf{v}_{B};\mathbf{v}_{C},%
\mathbf{v})  \label{f105} \\
&&\times \ \left[ f^{q}(\mathbf{v}_{A})\,f^{p}(\mathbf{v}_{B})-f^{q}(\mathbf{%
v}_{C})\,f^{p}(\mathbf{v})\right] \,d^{3}\mathbf{v}_{A}\,d^{3}\mathbf{v}%
_{B}\,d^{3}\mathbf{v}_{C}\,d^{3}\mathbf{v.}  \nonumber
\end{eqnarray}
\newline
From normalization of probabilities in eqs.(\ref{a30},\ref{f50}) we have
\begin{eqnarray}
\int G_{pq}^{pq}(\mathbf{v}_{A},\mathbf{v}_{B};\mathbf{v}_{C},\mathbf{v}%
)\,d^{3}\mathbf{v}_{C}\,d^{3}\mathbf{v} &=&1  \label{f111} \\
\int G_{pq}^{pq}(\mathbf{v}_{A},\mathbf{v}_{B};\mathbf{v}_{C},\mathbf{v}%
)\,d^{3}\mathbf{v}_{A}\,d^{3}\mathbf{v}_{B} &=&1.  \label{f111a}
\end{eqnarray}
Using these relations the integral on the right side of eq.(\ref{f105}) can
be written as
\begin{eqnarray}
\frac{d}{d\tau }\int f^{p}(\mathbf{v})d^{3}\mathbf{v} &=&\sum_{q=1}^{M}\int
f^{q}(\mathbf{v}_{A})\,\,f^{p}(\mathbf{v}_{B})\,d^{3}\mathbf{v}_{A}\,d^{3}%
\mathbf{v}_{B}  \label{f112} \\
&&-\sum_{q=1}^{M}\int f^{q}(\mathbf{v}_{C})\,\,f^{p}(\mathbf{v})d^{3}\mathbf{%
v}_{C}\,d^{3}\mathbf{v.}  \nonumber
\end{eqnarray}
These two terms are equal and they cancel each other yielding constancy of $%
\int f^{p}(\mathbf{v})d^{3}\mathbf{v}$.

The number of molecules of the $p^{th}\,$kind is
\begin{equation}
N_{p}=N\int f^{p}(\mathbf{v})\,d^{3}\mathbf{v,}  \label{f120}
\end{equation}
and it remains constant as it should. Hence the $F^{p}(\mathbf{v})$ is
normalized as
\begin{equation}
\int F^{p}(\mathbf{v})\,d^{3}\mathbf{v\,}d^{3}\mathbf{x=}N_{p},  \label{f125}
\end{equation}
where $\mathbf{x}$ is position of the molecule.

\subsection{One kind of gas with internal degrees of freedom}

For a homogeneous gas with internal states the $\mu $ stands for velocity $%
\mathbf{v}$ and a discrete index (for which we use $\alpha ,\beta ,i,j$)
defining the internal quantum state of the molecule. The mass of the
molecules is $m$. Particles with states $\mu _{A}=(\mathbf{v}_{A}\mathbf{,}%
\beta )$ and $\mu _{B}=(\mathbf{v}_{B}\mathbf{,}\alpha )$ enter the
collision and particles with states $\mu _{C}=(\mathbf{v}_{C}\mathbf{,}j)$
and $\mu _{D}=(\mathbf{v,}i)$ exits the collision. The integral over $\mu $
stands for integration over $\mathbf{v}$ and summation over the internal
state index. The internal energy of molecule in the state $\gamma $ is $%
E_{\gamma }$ and $\epsilon =E_{\alpha }+E_{\beta }$ and $\epsilon ^{\prime
}=E_{i}+E_{j}$. The center of mass (CM) coordinates are defined in eqs.(\ref
{e4},\ref{e5}).

Let us define the no time counter (NTC) kernel $K_{ij}^{\alpha \beta }(%
\mathbf{v}_{A},\mathbf{v}_{B};\mathbf{v}_{C},\mathbf{v})=K_{1}+K_{2}$ where
\begin{equation}
K_{1}=\frac{1}{R}\delta (\mathbf{H}-\mathbf{H}^{\prime })\,\delta \left[
\frac{2}{m_{r}}\epsilon +u^{2}-\frac{2}{m_{r}}\epsilon ^{\prime }-(u^{\prime
})^{2}\right] \,\frac{2u}{u^{\prime }}\,\,\sigma _{ij}^{\alpha \beta }(%
\mathbf{n},\mathbf{n}^{\prime }),  \label{g21}
\end{equation}
and
\begin{equation}
K_{2}=\left( 1-\frac{1}{R}\sum_{i}\sum_{j}u\Sigma _{ij}^{\alpha \beta
}\right) \delta (\mathbf{v}_{C}-\mathbf{v}_{A})\,\delta (\mathbf{v}-\mathbf{v%
}_{B})\,\delta _{i\alpha }\,\delta _{j\beta }.  \label{g30}
\end{equation}
Here $m_{r}=m/2$ is the reduced mass where $m$ is the mass of the molecules
and $R$ is a chosen parameter. The $\sigma _{ij}^{\alpha \beta }(\mathbf{n},%
\mathbf{n}^{\prime })$ is differential and the $\Sigma _{ij}^{\alpha \beta }$
is the total cross section into the internal states $i,j$
\begin{equation}
\Sigma _{ij}^{\alpha \beta }=\int \sigma _{ij}^{\alpha \beta }(\mathbf{n},%
\mathbf{n}^{\prime })\,d\mathbf{n}^{\prime },  \label{g40}
\end{equation}
where $d\mathbf{n}^{\prime }$ is the solid angle in the direction of $%
\mathbf{n}^{\prime }$. This kernel is symmetric due to the reciprocity
relation of the inelastic scattering cross sections\cite{Reciprocity}
\begin{equation}
u^{2}\,\sigma _{ij}^{\alpha \beta }(\mathbf{n},\mathbf{n}^{\prime
})=(u^{\prime })^{2}\,\sigma _{\alpha \beta }^{ij}(\mathbf{n}^{\prime },%
\mathbf{n}),  \label{g50}
\end{equation}
because $(u/u^{\prime })\,\sigma _{ij}^{\alpha \beta }=(u^{\prime
}/u)\,\sigma _{\alpha \beta }^{ij}$.

The $K_{2}$ part of $K_{ij}^{\alpha \beta }(\mathbf{v}_{A},\mathbf{v}_{B};%
\mathbf{v}_{C},\mathbf{v})$ directly transfers initial state to the final
state and causes a null collision. The probability of making a real
collision into the states $(i,j)$ is
\begin{equation}
P_{ij}=\int K_{1}\,d\mathbf{v}_{C}\,d\mathbf{v=}\frac{u\,\Sigma
_{ij}^{\alpha \beta }}{R}.  \label{g60}
\end{equation}
Therefore total probability of making a real collision is $%
(\sum_{i}\sum_{j}\,u\Sigma _{ij}^{\alpha \beta })/R$.

Inserting the $K_{ij}^{\alpha \beta }(\mathbf{v}_{A},\mathbf{v}_{B};\mathbf{v%
}_{C},\mathbf{v})$ into the eq.(\ref{a130}) and doing the integrals in the
CM\ coordinates we obtain
\begin{equation}
\frac{\partial f_{i}(\mathbf{v})}{\partial \tau }=\frac{1}{R}\sum_{\alpha
}\sum_{\beta }\sum_{j}\int \left[ f_{\beta }(\mathbf{v}_{A})\,f_{\alpha }(%
\mathbf{v}_{B})-f_{j}(\mathbf{v}_{C})\,f_{i}(\mathbf{v})\right] \,u\,\sigma
_{ij}^{\alpha \beta }(\mathbf{n},\mathbf{n}^{\prime })\,d^{3}\mathbf{u}\,d%
\mathbf{n}^{\prime }.  \label{g71}
\end{equation}
Here the $K_{2}$ part does not contribute to the collision integral as
before.

Again defining time as $t=\tau V/RN=2nV/RN^{2}$ and defining the new
functions $F_{i}(\mathbf{v})=(N/V)f_{i}(\mathbf{v})$ this is expressed as
\begin{equation}
\frac{\partial F_{i}}{\partial t}=\sum_{\alpha }\sum_{\beta }\sum_{j}\int
\left[ F_{\beta }(\mathbf{v}_{A})\,F_{\alpha }(\mathbf{v}_{B})-F_{j}(\mathbf{%
v}_{C})\,F_{i}(\mathbf{v})\right] \,\,u\,\sigma _{ij}^{\alpha \beta }(%
\mathbf{n},\mathbf{n}^{\prime })\,d^{3}\mathbf{u}\,d\mathbf{n}^{\prime },
\label{g80}
\end{equation}
where $\mathbf{v}_{A},\mathbf{v}_{B},\mathbf{v}_{C}$ are expressed in terms
of the variables $\mathbf{v,u,n}^{\prime }$ in eqs.(\ref{g75},\ref{g76},\ref
{g77},\ref{g78}). These equations are the Wang Chang-Uhlenbeck equations for
a gas with internal degrees of freedom. Here the states are assumed
nondegenerate for simplicity. Generalization to degenerate states is also
very straightforward.

Again we choose a number $R$ big enough such that for only very few (say
less than one in thousand) pairs $(\sum_{i}\sum_{j}u\Sigma _{ij}^{\alpha
\beta })/R$ will exceed unity. We chose $n=RN^{2}t/2V$ random pairs. For
each pair we take a random number $r$ and we allow the collision to happen
if $r<(\sum_{i}\sum_{j}u\Sigma _{ij}^{\alpha \beta })/R$. If collision is
allowed we choose the final state $(i,j)$ with the probability $\Sigma
_{ij}^{\alpha \beta }/(\sum_{i}\sum_{j}\Sigma _{ij}^{\alpha \beta })$ and
another random number is used to choose the final state. Finally we choose
the direction of scattering $\mathbf{n}^{\prime }$ according to the
probability density $\sigma _{ij}^{\alpha \beta }(\mathbf{n},\mathbf{n}%
^{\prime })/\Sigma _{ij}^{\alpha \beta }$ and a few more random numbers are
used for that. Then we calculate and store final velocities and state
indices for the colliding pair and go on to choose next pair.

\subsection{Mixture of gases with internal degrees of freedom}

This case is a combination of previous two cases and it is very
straightforward but unfortunately there are too many indices. The state of
particles are defined by three components of the velocity vector $\mathbf{v}$
and one kind index for which we use $p,q,r,s$ and one internal state index
for which we use $i,j,\alpha ,\beta $. We have $M$ kind of gas with internal
states in the mixture and there are $N_{p}$ number of molecules of the $%
p^{th}$ kind. The internal energy of $i^{th}$ internal state of $p^{th}$
kind molecule is $E_{i}^{p}$. The probability density $f(\mu )=f(\mathbf{v,}%
i,p)$ will be written as $f_{i}^{p}(\mathbf{v})$.

Particles with states $\mu _{A}=(\mathbf{v}_{A},\beta ,s),\mathbf{\ }$and $%
\mu _{B}=(\mathbf{v}_{B},\alpha ,r)$ enter the collision and particles with
states $\mu _{C}=(\mathbf{v}_{C},j,q)$ and $\mu _{D}=(\mathbf{v,}i,p\mathbf{)%
}$ exits the collision. We also define $\epsilon =E_{\beta }^{s}+E_{\alpha
}^{r}$ and $\epsilon ^{\prime }=E_{j}^{q}+E_{i}^{p}$. The integration over $%
\mu $ such as $\int f_{i}^{p}(\mathbf{v})d\mu $ stands for three
integrations over $\mathbf{v}$ and summations over $i$ and $p$. The center
of mass (CM) coordinates are defined in eqs.(\ref{e3},\ref{e3b},\ref{e4}).

The NTC kernel is $Q_{ij,pq}^{\alpha \beta ,rs}(\mathbf{v}_{A},\mathbf{v}%
_{B};\mathbf{v}_{C},\mathbf{v})=Q_{1}+Q_{2}$ where $Q_{1}$ and $Q_{2}$ are
defined as
\begin{equation}
Q_{1}=\frac{1}{R}\delta (\mathbf{H}-\mathbf{H}^{\prime })\,\delta \left[
\frac{2}{m_{r}}\epsilon +u^{2}-\frac{2}{m_{r}}\epsilon ^{\prime }-(u^{\prime
})^{2}\right] \frac{2u}{u^{\prime }}\,\,\sigma _{ij,pq}^{\alpha \beta ,pq}(%
\mathbf{n},\mathbf{n}^{\prime })\,\delta _{pr}\,\delta _{qs}.  \label{h21}
\end{equation}
and
\begin{equation}
Q_{2}=\left( 1-\frac{1}{R}\sum_{i}\sum_{j}u\Sigma _{ij,pq}^{\alpha \beta
,pq}\right) \delta (\mathbf{v}_{C}-\mathbf{v}_{A})\,\delta (\mathbf{v}-%
\mathbf{v}_{B})\,\delta _{i\alpha }\,\delta _{j\beta }\,\delta _{pr}\,\delta
_{qs}.  \label{h30}
\end{equation}
The delta functions $\delta _{pr}\delta _{qs}$ insures that the molecules do
no change identities during the collision. Here $%
m_{r}=m_{A}m_{B}/(m_{A}+m_{B})$ is the reduced mass, $R$ is a chosen
parameter. The $\sigma _{ij,pq}^{\alpha \beta ,pq}(\mathbf{n},\mathbf{n}%
^{\prime })$ is the differential cross section between species of the $%
p^{th} $ kind in the state $\alpha $ and $q^{th}$ kind in the state $\beta $
and $\Sigma _{ij,pq}^{\alpha \beta ,pq}$ is the total cross section into the
channel $(i,j)$
\begin{equation}
\Sigma _{ij,pq}^{\alpha \beta ,pq}=\int \sigma _{ij,pq}^{\alpha \beta ,pq}(%
\mathbf{n},\mathbf{n}^{\prime })\,d\mathbf{n}^{\prime }  \label{h40}
\end{equation}
where $d\mathbf{n}^{\prime }$ is the solid angle in the direction of $%
\mathbf{n}^{\prime }$. The $Q_{ij,pq}^{\alpha \beta ,rs}(\mathbf{v}_{A},%
\mathbf{v}_{B};\mathbf{v}_{C},\mathbf{v})$ is also symmetric due to eq.(\ref
{g50}). The $Q_{2}$ directly transfers initial states to the final states
and causes a null collision. The probability of making a real collision into
the states $(i,j)$ is
\begin{equation}
P_{ij}=\int Q_{1}\,d\mathbf{v}_{C}\,d\mathbf{v=}\frac{u\,\Sigma
_{ij,pq}^{\alpha \beta ,pq}}{R}  \label{h50}
\end{equation}
Therefore total probability of making a real collision is $%
(\sum_{i}\sum_{j}u\Sigma _{ij,pq}^{\alpha \beta ,pq})/R$.

Inserting the $Q_{ij,pq}^{\alpha \beta ,rs}(\mathbf{v}_{A},\mathbf{v}_{B};%
\mathbf{v}_{C},\mathbf{v})$ into the eq.(\ref{a130}) and doing the integrals
in the CM\ coordinates we obtain
\begin{equation}
\frac{\partial f_{i}^{p}(\mathbf{v})}{\partial \tau }=\sum_{q=1}^{M}\sum_{%
\alpha }\sum_{\beta }\sum_{j}\int [f^{q},f^{p}]_{ij}^{\alpha \beta
}\,Q_{ij,pq}^{\alpha \beta ,pq}(\mathbf{v}_{A},\mathbf{v}_{B};\mathbf{v}_{C},%
\mathbf{v})\,d^{3}\mathbf{v}_{A}\,d^{3}\mathbf{v}_{B}\,d^{3}\mathbf{v}_{C},
\label{h65}
\end{equation}
where
\begin{equation}
\lbrack f^{q},f^{p}]_{ij}^{\alpha \beta }=f_{\beta }^{q}(\mathbf{v}%
_{A})\,f_{\alpha }^{p}(\mathbf{v}_{B})-f_{j}^{q}(\mathbf{v}_{C})\,f_{i}^{p}(%
\mathbf{v}).  \label{h66}
\end{equation}
After inserting $Q_{ij,pq}^{\alpha \beta ,pq}$ we obtain
\begin{equation}
\frac{\partial f_{i}^{p}(\mathbf{v})}{\partial \tau }=\frac{1}{R}
\sum_{q=1}^{M}\sum_{\alpha }\sum_{\beta }\sum_{j}\int
[f^{q},f^{p}]_{ij}^{\alpha \beta }\,u\,\sigma _{ij,pq}^{\alpha \beta ,pq}(%
\mathbf{n},\mathbf{n}^{\prime })\,d^{3}\mathbf{u\,}d\mathbf{n}^{\prime }.
\label{h67}
\end{equation}
The $Q_{2}$ part does not contribute to the collision integral as before.
Expressions of $\mathbf{v}_{A},\mathbf{v}_{B},\mathbf{v}_{C}$ in terms of $%
\mathbf{v},\mathbf{u,}\mathbf{n}^{\prime }$ are given in eqs.(\ref{e100},\ref
{e101},\ref{e102},\ref{e103})

Again defining time as $t=\tau V/RN=2nV/RN^{2}$ and defining the new
functions $F_{i}^{p}(\mathbf{v})=(N/V)\,f_{i}^{p}(\mathbf{v})$ this is
expressed as
\begin{eqnarray}
\frac{\partial F_{i}^{p}(\mathbf{v})}{\partial t} &=&\sum_{q=1}^{M}\sum_{%
\alpha }\sum_{\beta }\sum_{j}\int \left( F_{\beta }^{q}(\mathbf{v}%
_{A})\,F_{\alpha }^{p}(\mathbf{v}_{B})-F_{j}^{q}(\mathbf{v}_{C})\,F_{i}^{p}(%
\mathbf{v})\right)  \label{h70} \\
&&\times u\,\sigma _{ij,pq}^{\alpha \beta ,pq}(\mathbf{n},\mathbf{n}^{\prime
})\,d^{3}\mathbf{u\,}d\mathbf{n}^{\prime }.  \nonumber
\end{eqnarray}
These equations are the Wang Chang-Uhlenbeck equations for a mixture of
gases with internal degrees of freedom. Here the states are assumed
nondegenerate for simplicity again.

Again we choose a number $R$ big enough such that for only very few (say
less than one in thousand) pairs $(\sum_{i}\sum_{j}u\Sigma _{ij,pq}^{\alpha
\beta ,pq})/R$ will exceed unity. We chose $n=RN^{2}t/2V$ random pairs. For
each pair we take a random number $r$ and we allow the collision to happen
if $r<(\sum_{i}\sum_{j}u\Sigma _{ij,pq}^{\alpha \beta ,pq})/R$. If collision
is allowed we choose the final state $(i,j)$ with the probability $\Sigma
_{ij,pq}^{\alpha \beta ,pq}/(\sum_{i}\sum_{j}\Sigma _{ij,pq}^{\alpha \beta
,pq})$ and another random number is used to choose the final state. Finally
we choose the direction of scattering $\mathbf{n}^{\prime }$ according to
the probability density $\sigma _{ij,pq}^{\alpha \beta ,pq}(\mathbf{n},%
\mathbf{n}^{\prime })/\Sigma _{ij,pq}^{\alpha \beta ,pq}$ and a few more
random numbers are used for that. Then we calculate and store final
velocities and state indices for the colliding pair and go on to choose next
pair.

Note that the normalization of $f_{i}^{p}(\mathbf{v})\,$is given by
\begin{equation}
\sum_{p}\sum_{i}\int f_{i}^{p}(\mathbf{v})\,d^{3}\mathbf{v}=1.  \label{h80}
\end{equation}
The expression $\sum_{i}\int f_{i}^{p}(\mathbf{v})d^{3}\mathbf{v}$ is
conserved during the simulation. From eq.(\ref{h65}) its rate of change is
\begin{eqnarray}
\frac{d}{d\tau }\sum_{i}\int f_{i}^{p}(\mathbf{v})d^{3}\mathbf{v}
&=&\sum_{i}\int \frac{\partial f_{i}^{p}(\mathbf{v})}{\partial \tau }\,d^{3}%
\mathbf{v}=\sum_{q=1}^{M}\sum_{\alpha }\sum_{\beta }\sum_{i}\sum_{j}
\label{h91} \\
&&\int [f^{q},f^{p}]_{ij}^{\alpha \beta }\,Q_{ij,pq}^{\alpha \beta ,pq}(%
\mathbf{v}_{A},\mathbf{v}_{B};\mathbf{v}_{C},\mathbf{v})\,d^{3}\mathbf{v}%
_{A}\,d^{3}\mathbf{v}_{B}\,d^{3}\mathbf{v}_{C}\,d^{3}\mathbf{v}  \nonumber
\end{eqnarray}
\newline
From symmetry and normalization of the kernel given in eqs.(\ref{a10},\ref
{a20},\ref{a30}) we have
\begin{eqnarray}
\sum_{i}\sum_{j}\int Q_{ij,pq}^{\alpha \beta ,pq}(\mathbf{v}_{A},\mathbf{v}%
_{B};\mathbf{v}_{C},\mathbf{v})\,d^{3}\mathbf{v}_{C}\,d^{3}\mathbf{v} &=&1
\label{h120} \\
\sum_{\alpha }\sum_{\beta }\int Q_{ij,pq}^{\alpha \beta ,pq}(\mathbf{v}_{A},%
\mathbf{v}_{B};\mathbf{v}_{C},\mathbf{v})\,d^{3}\mathbf{v}_{A}\,d^{3}\mathbf{%
v}_{B} &=&1  \label{h121}
\end{eqnarray}
Using this, we express eq.(\ref{h91}) as
\begin{eqnarray}
\frac{d}{d\tau }\sum_{i}\int f_{i}^{p}(\mathbf{v})\,d^{3}\mathbf{v}
&=&\sum_{q=1}^{M}\sum_{\alpha }\sum_{\beta }\int f_{\beta }^{q}(\mathbf{v}%
_{A})\,f_{\alpha }^{p}(\mathbf{v}_{B})\,d^{3}\mathbf{v}_{A}\,d^{3}\mathbf{v}%
_{B}  \label{h130} \\
&&-\sum_{q=1}^{M}\sum_{i}\sum_{j}\int f_{j}^{q}(\mathbf{v}_{C})\,f_{i}^{p}(%
\mathbf{v})\,d^{3}\mathbf{v}_{C}\,d^{3}\mathbf{v}  \nonumber
\end{eqnarray}
These two terms are equal and they cancel each other yielding constancy of $%
\sum_{i}f_{i}^{p}(\mathbf{v})\,d^{3}\mathbf{v}$. The number of molecules of
the $p^{th}\,$kind is
\begin{equation}
N_{p}=N\sum_{i}\int f_{i}^{p}(\mathbf{v})\,d^{3}\mathbf{v}  \label{h140}
\end{equation}
and as the above argument shows, it remains constant as it should. Hence the
$F_{i}^{p}(\mathbf{v})$ is normalized as
\begin{equation}
\sum_{i}\int F_{i}^{p}(\mathbf{v})\,d^{3}\mathbf{v\,}d^{3}\mathbf{x=}N_{p},
\label{h150}
\end{equation}
where $\mathbf{x}$ is position of the molecule.

\subsection{Relation to Kac's work}

Fifty years ago M. Kac\cite{Kac} introduced a master equation similar to
ours and derived the Boltzmann equation for a homogenous gas from it. Here
we summarize his work and point out similarities. We will use a different
notation than his.

Suppose we have $N$ particles in a gas contained in volume $V$. Collisions
are assumed to take place randomly within the gas. Again we have a
probability distribution $f^{(N)}(\mathbf{v}_{1},\mathbf{v}_{2},...,\mathbf{v%
}_{N};t)$ for their velocities. For brevity we will show this as $f^{(N)}(%
\mathbf{v};t)$ wherever convenient. Probability that the $i^{th}$ and $%
j^{th} $ particles having velocities $\mathbf{v}_{A}$ and $\mathbf{v}_{B}$
will collide and emerge with velocities $\mathbf{v}_{C}\,$and $\mathbf{v}%
_{D} $ in the phase space $d^{3}\mathbf{v}_{C}d^{3}\mathbf{v}_{D}$ in a time
interval $dt$ is $R(\mathbf{v}_{A},\mathbf{v}_{B};\mathbf{v}_{C},\mathbf{v}%
_{D})d^{3}\mathbf{v}_{C}d^{3}\mathbf{v}_{D}dt.$ Here the $R(\mathbf{v}_{A},%
\mathbf{v}_{B};\mathbf{v}_{C},\mathbf{v}_{D})$ is a function connected to
differential cross section but we will not need the precise relation until
later. The total collision probability in $dt$ time interval is $S(\mathbf{v}%
_{A},\mathbf{v}_{B})dt$ where $S(\mathbf{v}_{A},\mathbf{v}_{B})\,$is given
by
\begin{equation}
S(\mathbf{v}_{A},\mathbf{v}_{B})=\int R(\mathbf{v}_{A},\mathbf{v}_{B};%
\mathbf{v}_{C},\mathbf{v}_{D})d^{3}\mathbf{v}_{C}d^{3}\mathbf{v}_{D}.
\label{kac10}
\end{equation}
As usual we assume some symmetries for the $R(\mathbf{v}_{A},\mathbf{v}_{B},%
\mathbf{v}_{C},\mathbf{v}_{D})$ function:
\begin{eqnarray}
R(\mathbf{v}_{A},\mathbf{v}_{B};\mathbf{v}_{C},\mathbf{v}_{D}) &=&R(\mathbf{v%
}_{C},\mathbf{v}_{D};\mathbf{v}_{A},\mathbf{v}_{B}),  \label{kac20} \\
R(\mathbf{v}_{A},\mathbf{v}_{B};\mathbf{v}_{C},\mathbf{v}_{D}) &=&R(\mathbf{v%
}_{B},\mathbf{v}_{A};\mathbf{v}_{D},\mathbf{v}_{C}).  \label{kac30}
\end{eqnarray}

The $f^{(N)}(\mathbf{v}_{1},\mathbf{v}_{2},...,\mathbf{v}_{N};t)$ satisfies
the master equation
\begin{equation}
\frac{\partial f^{(N)}(\mathbf{v})}{\partial t}=-f^{(N)}(\mathbf{v}%
)\sum_{i=1}^{N}\sum_{j\neq i}^{N}S(\mathbf{v}_{i},\mathbf{v}%
_{j})+\sum_{i=1}^{N}\sum_{j\neq i}^{N}\int f_{ij}^{(N)}(\mathbf{v}_{A},%
\mathbf{v}_{B})R(\mathbf{v}_{A},\mathbf{v}_{B};\mathbf{v}_{i},\mathbf{v}%
_{j})d^{3}\mathbf{v}_{A}d^{3}\mathbf{v}_{B}  \label{kac40}
\end{equation}
In order to see where this comes from we write it for infinitesimal time
interval $dt$:
\begin{eqnarray}
f^{(N)}(\mathbf{v;}t+dt) &=&f^{(N)}(\mathbf{v;}t)\left(
1-dt\sum_{i=1}^{N}\sum_{j\neq i}^{N}S(\mathbf{v}_{i},\mathbf{v}_{j})\right)
\label{kac50} \\
&&+dt\left( \sum_{i=1}^{N}\sum_{j\neq i}^{N}\int f_{ij}^{(N)}(\mathbf{v}_{A},%
\mathbf{v}_{B})R(\mathbf{v}_{A},\mathbf{v}_{B};\mathbf{v}_{i},\mathbf{v}%
_{j})d^{3}\mathbf{v}_{A}d^{3}\mathbf{v}_{B}\right) .  \nonumber
\end{eqnarray}
Let us multiply both sides with $d^{3}\mathbf{v}_{1}...d^{3}\mathbf{v}_{N}.$
Then $f^{(N)}(\mathbf{v;}t+dt)d^{3}\mathbf{v}_{1}...d^{3}\mathbf{v}_{N}$ is
the probability that the velocities are in the phase space volume $d^{3}%
\mathbf{v}_{1}...d^{3}\mathbf{v}_{N}$ at time $t+dt$. The first term on the
right is
\begin{equation}
\left( f^{(N)}(\mathbf{v;}t)d^{3}\mathbf{v}_{1}...d^{3}\mathbf{v}_{N}\right)
\left( 1-dt\sum_{i=1}^{N}\sum_{j\neq i}^{N}S(\mathbf{v}_{i},\mathbf{v}%
_{j})\right) .  \label{kac60}
\end{equation}
The first parenthesis is the probability that the system was in $d^{3}%
\mathbf{v}_{1}...d^{3}\mathbf{v}_{N}$ phase space volume at time $t$ and the
second parenthesis is the probability that no collisions occurred in $dt$
time interval. Their product is the probability of arriving $d^{3}\mathbf{v}%
_{1}...d^{3}\mathbf{v}_{N}$ phase space volume at $t+dt$ without making a
collision. The second term in the right side are probabilities of arriving
in $d^{3}\mathbf{v}_{1}...d^{3}\mathbf{v}_{N}$ by making collisions with
different pairs. For example let us write $i=1,$ $j=2$ term:
\begin{equation}
\int \left( f^{(N)}(\mathbf{v}_{A},\mathbf{v}_{B},\mathbf{v}_{3},...,\mathbf{%
v}_{N})d^{3}\mathbf{v}_{A}d^{3}\mathbf{v}_{B}d^{3}\mathbf{v}_{3}...d^{3}%
\mathbf{v}_{N}\right) \left( R(\mathbf{v}_{A},\mathbf{v}_{B};\mathbf{v}_{1},%
\mathbf{v}_{2})d^{3}\mathbf{v}_{1}d^{3}\mathbf{v}_{2}dt\right) .
\label{kac70}
\end{equation}
The first parenthesis under the integral is the probability that the system
was in the phase space volume $d^{3}\mathbf{v}_{A}d^{3}\mathbf{v}_{B}d^{3}%
\mathbf{v}_{3}...d^{3}\mathbf{v}_{N}$ at time $t$ and the second parenthesis
is the probability that the collision between particles one and two took
them to $d^{3}\mathbf{v}_{1}d^{3}\mathbf{v}_{2}$ phase space volume. If we
integrate this product over $\mathbf{v}_{A},\mathbf{v}_{B}$ we obtain
probability of arriving in $d^{3}\mathbf{v}_{1}...d^{3}\mathbf{v}_{N}$ at
time $t+dt$ via a collision between particles one and two. To obtain total
probability of arriving in $d^{3}\mathbf{v}_{1}...d^{3}\mathbf{v}_{N}$ at
time $t+dt$ via a collision we sum such terms over all possible pairs. This
argument clearly shows how the master equation is derived.

Writing $S(\mathbf{v}_{i},\mathbf{v}_{j})$ as
\begin{equation}
S(\mathbf{v}_{i},\mathbf{v}_{j})=\int R(\mathbf{v}_{A},\mathbf{v}_{B};%
\mathbf{v}_{j},\mathbf{v}_{i})d^{3}\mathbf{v}_{A}d^{3}\mathbf{v}_{B}
\label{kac80}
\end{equation}
the master equation can be written in a more symmetric form
\begin{equation}
\frac{\partial f^{(N)}(\mathbf{v})}{\partial t}=\sum_{i=1}^{N}\sum_{j\neq
i}^{N}\int \left( f_{ij}^{(N)}(\mathbf{v}_{A},\mathbf{v}_{B})-f^{(N)}(%
\mathbf{v})\right) R(\mathbf{v}_{A},\mathbf{v}_{B};\mathbf{v}_{i},\mathbf{v}%
_{j})d^{3}\mathbf{v}_{A}d^{3}\mathbf{v}_{B}  \label{kac90}
\end{equation}

All of the results we obtained from our master equation can be obtained for
this master equation too. Kac\cite{Kac} showed that the distribution goes to
microcanonical distribution as $t\rightarrow \infty $. A hierarchy of
reduced probability equations can be obtained for this master equation too.
Kac\cite{Kac} showed that in the limit $N\rightarrow \infty $ if one starts
from uncorrelated state at $t=0$ the system always remains uncorrelated. His
arguments was different than ours.

The first equation in the hierarchy (obtained by integrating over $\mathbf{v}%
_{2},\mathbf{v}_{3},...,\mathbf{v}_{N}$ ) is
\begin{equation}
\frac{\partial f^{(1)}(\mathbf{v})}{\partial t}=2N\int \left( f^{(2)}(%
\mathbf{v}_{A},\mathbf{v}_{B})-f^{(2)}(\mathbf{v,v}_{C})\right) R(\mathbf{v}%
_{A},\mathbf{v}_{B};\mathbf{v}_{C},\mathbf{v})d^{3}\mathbf{v}_{A}d^{3}%
\mathbf{v}_{B}d^{3}\mathbf{v}_{C}  \label{kac100}
\end{equation}
If we introduce AMC this equation becomes
\begin{equation}
\frac{\partial f(\mathbf{v})}{\partial t}=2N\int \left( f(\mathbf{v}_{A})f(%
\mathbf{v}_{B})-f(\mathbf{v})f(\mathbf{v}_{C})\right) R(\mathbf{v}_{A},%
\mathbf{v}_{B};\mathbf{v}_{C},\mathbf{v})d^{3}\mathbf{v}_{A}d^{3}\mathbf{v}%
_{B}d^{3}\mathbf{v}_{C}.  \label{kac110}
\end{equation}
Here the superscript (1) is dropped and time $t$ is suppressed in $f^{(1)}(%
\mathbf{v;}t)$.

Now we go to center of mass frame (Equations \ref{e4},\ref{e5}). In the CM
coordinates the $R(\mathbf{v}_{A},\mathbf{v}_{B};\mathbf{v}_{C},\mathbf{v})$
is expressed as
\begin{equation}
R(\mathbf{v}_{A},\mathbf{v}_{B};\mathbf{v}_{C},\mathbf{v})=\frac{1}{V}\delta
\left( \mathbf{H}-\mathbf{H}^{\prime }\right) \,\delta \left(
u^{2}-(u^{\prime })^{2}\right) \,\sigma (\mathbf{n},\mathbf{n}^{\prime })
\label{kac140}
\end{equation}
where $V$ is the volume of the gas and $\sigma (\mathbf{n},\mathbf{n}%
^{\prime })$ is the differential cross section. Inserting this into eq.(\ref
{kac110}) and doing the integrals over the center of mass frame we obtain
\begin{equation}
\frac{\partial f(\mathbf{v})}{\partial t}=\frac{N}{V}\int \left[ f(\mathbf{v}%
_{A})\,f(\mathbf{v}_{B})-f(\mathbf{v}_{C})\,f(\mathbf{v})\right] \,u\,\sigma
(\mathbf{n},\mathbf{n}^{\prime })\,d^{3}\mathbf{u}\,d\mathbf{n}^{\prime },
\label{kac150}
\end{equation}
where $\mathbf{v}_{A},\mathbf{v}_{B},\mathbf{v}_{C}$ are expressed in terms
of the variables $\mathbf{v,u,n}^{\prime }$ in eqs.(\ref{e105},\ref{e106},%
\ref{e107}). If we write this equation for $F(\mathbf{v})=(N/V)f(\mathbf{v})$
which is velocity distribution normalized to the number density per unit
volume, we obtain the Boltzmann equation for a homogenous gas
\begin{equation}
\frac{\partial F(\mathbf{v})}{\partial t}=\int \left[ F(\mathbf{v}_{A})\,F(%
\mathbf{v}_{B})-F(\mathbf{v}_{C})\,F(\mathbf{v})\right] \,u\,\sigma (\mathbf{%
n},\mathbf{n}^{\prime })\,d^{3}\mathbf{u}\,d\mathbf{n}^{\prime }.
\label{kac160}
\end{equation}

Although both master equations have similar structures their philosophies
are different. In Kac's work the collisions happens randomly and
spontaneously in the gas whereas in direct simulation we take pairs and
force them to collide. Direct simulation has applications to systems other
than gases as we showed in the money games examples. In these systems there
are not physical processes driving the collisions and instead we make the
collisions. In Kac's work his motivation was to describe Boltzmann equation
for gases as a stochastic equation and the DSMC method had not been invented
yet. Just as in our work, Kac's method can be generalized to molecular gases
and gas mixtures and one can obtain Boltzmann equations for these cases by
defining a suitable $R(\mathbf{v}_{A},\mathbf{v}_{B};\mathbf{v}_{C},\mathbf{v%
})$ for each case.

\section{Direct simulation for an inhomogeneous gas}

In this section we study NTC algorithm of DSMC method for inhomogeneous gas.
We will not actually derive Bird's algorithm but we will define a similar
algorithm to simulate inhomogeneous gas. We will show that single particle
probability distribution of our algorithm satisfies the Boltzmann equation
for an inhomogeneous gas. Then we will argue that both algorithms give the
same results in the limit $N\rightarrow \infty $.

We divide the physical space into cells as in the Bird's method. In our
algorithm we take pairs not from the same cell but from all of the volume
and we let each pair to make a collision attempt if both of them are in the
same cell.

We divide the physical space into cells and the $k^{th}$ cell has the volume
$V_{k}$. Now let us define the functions
\begin{equation}
\Delta _{k}(\mathbf{x})=\left\{
\begin{array}{ll}
1\quad & \mathbf{{}x}\in V_{k} \\
0\quad & \mathbf{x}\notin V_{k}
\end{array}
\right. .  \label{k10}
\end{equation}
We will also need the function
\begin{equation}
\Gamma (\mathbf{x},\mathbf{x}^{\prime })=\sum_{k}\frac{\Delta _{k}(\mathbf{x}%
)\Delta _{k}(\mathbf{x}^{\prime })}{V_{k}}.  \label{k20}
\end{equation}
This function is zero when $\mathbf{x}$ and $\mathbf{x}^{\prime }$ are not
in the same cell and $1/V_{k}$ when they are in the same cell. Its integral
over $\mathbf{x}$ or $\mathbf{x}^{\prime }$ is unity
\begin{equation}
\int \Gamma (\mathbf{x},\mathbf{x}^{\prime })d^{3}\mathbf{x}^{\prime }=\int
\Gamma (\mathbf{x},\mathbf{x}^{\prime })d^{3}\mathbf{x}=1.  \label{k30}
\end{equation}
At the end of this section we will take the limit $V_{k}\rightarrow 0$. In
this limit $\Gamma (\mathbf{x},\mathbf{x}^{\prime })=0$ for $\mathbf{x}\neq
\mathbf{x}^{\prime }$ and $\Gamma (\mathbf{x},\mathbf{x}^{\prime })=\infty $
for $\mathbf{x}=\mathbf{x}^{\prime }$ and eq.(\ref{k30}) is still satisfied.
These are properties of the Dirac delta function and we have the limit
\begin{equation}
\lim_{V_{k}\rightarrow 0}\Gamma (\mathbf{x},\mathbf{x}^{\prime })=\delta (%
\mathbf{x}-\mathbf{x}^{\prime })  \label{k40}
\end{equation}

Now we can start the discussion. We will treat the simplest case for
clarity. We develop our arguments for one kind of gas without internal
degrees of freedom. The generalization to the other cases is very
straightforward and will be summarized at the end of the section.

The state index $\mu $ represent position of the particle $\mathbf{x}$ and
the velocity $\mathbf{v.}$ The collision kernel is $Z=Z_{1}+Z_{2}$ where $%
Z_{1}$ and $Z_{2}$ are
\begin{eqnarray}
Z_{1}(\mathbf{x}_{A}\mathbf{v}_{A},\mathbf{x}_{B}\mathbf{v}_{B};\mathbf{x}%
_{C}\mathbf{v}_{C},\mathbf{x}_{D}\mathbf{v}_{D}) &=&S(\mathbf{v}_{A},\mathbf{%
v}_{B};\mathbf{v}_{C}\mathbf{,v}_{D})\,  \label{k50} \\
&&\times \Gamma (\mathbf{x}_{A},\mathbf{x}_{B})\,\Omega \,\delta (\mathbf{x}%
_{C}-\mathbf{x}_{A})\,\delta (\mathbf{x}_{D}-\mathbf{x}_{B})\,,  \nonumber
\end{eqnarray}
and
\begin{eqnarray}
Z_{2}(\mathbf{x}_{A}\mathbf{v}_{A},\mathbf{x}_{B}\mathbf{v}_{B};\mathbf{x}%
_{C}\mathbf{v}_{C},\mathbf{x}_{D}\mathbf{v}_{D}) &=&\,\,\left( 1-\Omega
\,\Gamma (\mathbf{x}_{A},\mathbf{x}_{B})\right) \delta (\mathbf{x}_{C}-%
\mathbf{x}_{A})\,\,\,  \label{k61} \\
&&\,\,\times \delta (\mathbf{x}_{D}-\mathbf{x}_{B})\,\delta (\mathbf{v}_{C}-%
\mathbf{v}_{A})\,\delta (\mathbf{v}_{D}-\mathbf{v}_{B}).  \nonumber
\end{eqnarray}
Here
\begin{equation}
\Omega \,=\left( \sum_{k}\frac{1}{V_{k}}\right) ^{-1},  \label{k70}
\end{equation}
is a constant chosen to insure that probability of making a collision in any
cell is less than unity. The $S(\mathbf{v}_{A},\mathbf{v}_{B};\mathbf{v}_{C}%
\mathbf{,v}_{D})$ is given in eqs.(\ref{e20},\ref{e30}). The $Z_{2}$ does
not change states of the of the particles and the pair will not be allowed
to make a collision attempt with a probability $\left( 1-\Omega \,\Gamma (%
\mathbf{x}_{A},\mathbf{x}_{B})\right) .$ The probability of a collision
attempt is $\Omega \Gamma (\mathbf{x}_{A},\mathbf{x}_{B})$ and in a real
collision positions of particles do not change because of the $\delta (%
\mathbf{x}_{C}-\mathbf{x}_{A})\,\delta (\mathbf{x}_{D}-\mathbf{x}_{B})$ term
in the $Z$. The $Z(\mathbf{x}_{A}\mathbf{v}_{A},\mathbf{x}_{B}\mathbf{v}_{B};%
\mathbf{x}_{C}\mathbf{v}_{C},\mathbf{x}_{D}\mathbf{v}_{D})$ is symmetric and
satisfies the normalization condition
\begin{eqnarray}
\int Z(\mathbf{x}_{A}\mathbf{v}_{A},\mathbf{x}_{B}\mathbf{v}_{B};\mathbf{x}%
_{C}\mathbf{v}_{C},\mathbf{x}_{D}\mathbf{v}_{D})d^{3}\mathbf{v}_{A}d^{3}%
\mathbf{v}_{B}d^{3}\mathbf{x}_{A}d^{3}\mathbf{x}_{B} &=&1,  \label{k81} \\
\int Z(\mathbf{x}_{A}\mathbf{v}_{A},\mathbf{x}_{B}\mathbf{v}_{B};\mathbf{x}%
_{C}\mathbf{v}_{C},\mathbf{x}_{D}\mathbf{v}_{D})d^{3}\mathbf{v}_{C}d^{3}%
\mathbf{v}_{D}d^{3}\mathbf{x}_{C}d^{3}\mathbf{x}_{D} &=&1.  \label{k82}
\end{eqnarray}

The $\Omega \Gamma (\mathbf{x}_{A},\mathbf{x}_{B})$ vanishes unless $\mathbf{%
x}_{A}$ and $\mathbf{x}_{B}$ are in the same cell and $\Omega \Gamma (%
\mathbf{x}_{A},\mathbf{x}_{B})=\Omega /V_{k}$ when $\mathbf{x}_{A}$ and $%
\mathbf{x}_{B}$ are in the cell $V_{k}$. The probability of having both
particles in the cell $V_{k}$ is $(N_{k}/N)^{2}$ where $N_{k}$ is the number
of particles in the cell $V_{k}$ during the collisions part of the
simulation. Therefore the probability of a pair making a collision attempt
in the $k^{th}$ cell is $p_{k}=(\Omega /V_{k})(N_{k}/N)^{2}$. The $1/V_{k}\,$%
term looks awkward in this probability but it is absolutely necessary as the
following argument shows. Suppose the physical density is uniform and
therefore $N_{k}/N=V_{k}/V$ where $V$ is the total volume. When density is
uniform we expect that the probability of having a collision in $V_{k}$ is
proportional to $V_{k}.$ When $N_{k}/N=V_{k}/V$ in inserted in $p_{k}\,$we
find $p_{k}=\Omega V_{k}/V^{2}$ which is proportional to $V_{k}$ as expected.

Now we insert the kernel $Z\,$in the eq.(\ref{a130}) to obtain
\begin{eqnarray}
\frac{\partial f(\mathbf{xv},\tau \mathbf{)}}{\partial \tau } &=&\int
[f,f]\,Z(\mathbf{x}_{A}\mathbf{v}_{A},\mathbf{x}_{B}\mathbf{v}_{B};\mathbf{x}%
_{C}\mathbf{v}_{C},\mathbf{xv})  \label{k90} \\
&&\times d^{3}\mathbf{v}_{A}\,d^{3}\mathbf{v}_{B}\,d^{3}\mathbf{v}_{C}\,d^{3}%
\mathbf{x}_{A}\,d^{3}\mathbf{x}_{B}\,d^{3}\mathbf{x}_{C}  \nonumber
\end{eqnarray}
where $[f,f]\,$is
\begin{equation}
\lbrack f,f]\,=f(\mathbf{x}_{A}\mathbf{v}_{A},\tau )f(\mathbf{x}_{B}\mathbf{v%
}_{B},\tau )-f(\mathbf{x}_{C}\mathbf{v}_{C},\tau )f(\mathbf{xv},\tau ).
\label{k100}
\end{equation}
The $Z_{2}$ part of the collision kernel does not contribute to the
collision integral. After doing the delta function integrals over positions $%
\mathbf{x}_{A}\,,\mathbf{x}_{B}$ we obtain
\begin{eqnarray}
\frac{\partial f(\mathbf{xv},\tau \mathbf{)}}{\partial \tau } &=&\Omega \int
\left[ f(\mathbf{x}^{\prime }\mathbf{v}_{A},\tau )f(\mathbf{xv}_{B},\tau )-f(%
\mathbf{x}^{\prime }\mathbf{v}_{C},\tau )f(\mathbf{xv},\tau )\right]
\label{k110} \\
&&\times \Gamma (\mathbf{x},\mathbf{x}^{\prime })\,S(\mathbf{v}_{A},\mathbf{v%
}_{B};\mathbf{v}_{C}\mathbf{,v})d^{3}\mathbf{v}_{A}\,d^{3}\mathbf{v}%
_{B}\,d^{3}\mathbf{v}_{C}\,d^{3}\mathbf{x}^{\prime }.  \nonumber
\end{eqnarray}
Now we insert $S=S_{1}+S_{2}$ from eqs.(\ref{e20},\ref{e30}) in this
equation. The $S_{2}$ part gives no contribution to the integral as before.
Doing the integrals over $\mathbf{v}_{A}\,,\mathbf{v}_{B}\,,\mathbf{v}_{C}\,$
in the center of mass coordinates we obtain
\begin{eqnarray}
\frac{\partial f(\mathbf{xv},\tau \mathbf{)}}{\partial \tau } &=&\frac{%
\Omega }{R}\int \left[ f(\mathbf{x}^{\prime }\mathbf{v}_{A},\tau )f(\mathbf{%
xv}_{B},\tau )-f(\mathbf{x}^{\prime }\mathbf{v}_{C},\tau )f(\mathbf{xv},\tau
)\right]  \label{k113} \\
&&\times \Gamma (\mathbf{x},\mathbf{x}^{\prime })\,\sigma (\mathbf{n,n}%
^{\prime })\,d^{3}\mathbf{u}\,d\mathbf{n}^{\prime }\,d^{3}\mathbf{x}^{\prime
}.  \nonumber
\end{eqnarray}
where $\mathbf{v}_{A},\mathbf{v}_{B},\mathbf{v}_{C}$ are given in eqs.(\ref
{e105},\ref{e106},\ref{e107}). In order to have complete correspondence with
the Boltzmann equation we define the new function $F(\mathbf{xv},\tau )=Nf(%
\mathbf{xv},\tau \mathbf{)}$ and we also define the new variable $t=\Omega
\tau /RN=2\Omega n/RN^{2}$ to obtain
\begin{equation}
\frac{\partial F(\mathbf{xv},t\mathbf{)}}{\partial t}=\widehat{L}_{C}F(%
\mathbf{xv,}t\mathbf{)}  \label{k122}
\end{equation}

where the operator $\widehat{L}_{C}$ is defined as
\begin{eqnarray}
\widehat{L}_{C}F(\mathbf{xv,}t\mathbf{)} &=&\ \int \left[ F(\mathbf{x}%
^{\prime }\mathbf{v}_{A},t)F(\mathbf{xv}_{B},t)-F(\mathbf{x}^{\prime }%
\mathbf{v}_{C},t)F(\mathbf{xv},t)\right]  \label{k150} \\
&&\times \Gamma (\mathbf{x},\mathbf{x}^{\prime })\,\sigma (\mathbf{n,n}%
^{\prime })\,d^{3}\mathbf{u}\,d\mathbf{n}^{\prime }\,d^{3}\mathbf{x}^{\prime
}.  \nonumber
\end{eqnarray}
Here $t$ is interpreted as the physical time.

In the collisions part of the DSMC\ method we make collision attempts for a
time $\Delta t$ where $\Delta t$ is a small time interval. This corresponds
to $\Delta \tau =RN\Delta t/\Omega $ collision time passage or $\Delta
n=RN^{2}\Delta t/2\Omega $ pairs chosen. From eq.(\ref{k122}), after making $%
\Delta n$ collisions attempt $F(\mathbf{xv},t\mathbf{)}$ becomes $F^{*}(%
\mathbf{xv},t\mathbf{)}$%
\begin{equation}
F^{*}(\mathbf{xv},t\mathbf{)=}(1+\Delta t\widehat{L}_{C})F(\mathbf{xv},t%
\mathbf{)+}O((\Delta t)^{2})  \label{m10}
\end{equation}
where $O((\Delta t)^{2})$ is an error term of order $(\Delta t)^{2}$.

Next we perform free propagation step where $\mathbf{x\rightarrow x+}\Delta t%
\mathbf{v}$ and $\mathbf{v\rightarrow v+}\Delta t\mathbf{a}$ transformation
is made for each particle. Here $\mathbf{a=F/}m$ is the acceleration of the
particle due to the force $\mathbf{F}$ and it can depend on both position
and velocity of the particle. This changes the $N$ particle distribution
function $f^{(N)}(\mathbf{x}_{1},\mathbf{v}_{1};\mathbf{x}_{2},\mathbf{v}%
_{2};...,\mathbf{x}_{N},\mathbf{v}_{N})$ to
\begin{equation}
f^{(N)}(\mathbf{x}_{1}-\Delta t\mathbf{v}_{1},\mathbf{v}_{1}-\Delta t\mathbf{%
a}_{1};...;\mathbf{x}_{N}-\Delta t\mathbf{v}_{N},\mathbf{v}_{N}-\Delta t%
\mathbf{a}_{N}).  \label{m20}
\end{equation}
The jacobian of the transformation is unity with a correction of order $%
(\Delta t)^{2}$ and therefore this expression is correct with an error of
the same order. Integrating this over $\mathbf{x}_{2},\mathbf{v}_{2};...,%
\mathbf{x}_{N},\mathbf{v}_{N}$ we find that the single particle probability
distribution $f^{(1)}(\mathbf{x},\mathbf{v)}$ changes to $f^{(1)}(\mathbf{x-}%
\Delta t\mathbf{v},\mathbf{v-}\Delta t\mathbf{a)}$ with an error term of
order $(\Delta t)^{2}$. Therefore $F^{*}(\mathbf{x},\mathbf{v},t\mathbf{)}$
becomes $F^{*}(\mathbf{x-}\Delta t\mathbf{v},\mathbf{v-}\Delta t\mathbf{a,}t%
\mathbf{)}$ which is taken as $F(\mathbf{x},\mathbf{v},t+\Delta t\mathbf{)}$%
. Hence
\begin{equation}
F(\mathbf{x},\mathbf{v},t+\Delta t\mathbf{)}=F^{*}(\mathbf{x-}\Delta t%
\mathbf{v},\mathbf{v-}\Delta t\mathbf{a,}t\mathbf{).}  \label{m31}
\end{equation}
Using eq.(\ref{m10}) and expanding $F(\mathbf{x-}\Delta t\mathbf{v},\mathbf{%
v-}\Delta t\mathbf{a,}t\mathbf{)}$ up to first order terms in $\Delta t$ we
obtain
\begin{equation}
F(\mathbf{x},\mathbf{v},t+\Delta t\mathbf{)}=\left( 1-\Delta t\mathbf{v}%
\frac{\partial }{\partial \mathbf{x}}-\Delta t\mathbf{a}\frac{\partial }{%
\partial \mathbf{v}}+\Delta t\widehat{L}_{C}\right) F(\mathbf{x},\mathbf{v}%
,t)+O((\Delta t)^{2})  \label{m35}
\end{equation}
where $O((\Delta t)^{2})$ is the error terms of order $(\Delta t)^{2}$.
Taking the limit $\Delta t\rightarrow 0$ we obtain
\begin{equation}
\frac{\partial F(\mathbf{x},\mathbf{v},t\mathbf{)}}{\partial t}+\mathbf{%
v\cdot }\frac{\partial F(\mathbf{x},\mathbf{v},t\mathbf{)}}{\partial \mathbf{%
x}}+\frac{\mathbf{F}}{m}\mathbf{\cdot }\frac{\partial F(\mathbf{x},\mathbf{v}%
,t\mathbf{)}}{\partial \mathbf{v}}=\widehat{L}_{C}F(\mathbf{x},\mathbf{v},t%
\mathbf{)}  \label{m40}
\end{equation}

This equation is similar to the Boltzmann equation but it is not the same.
Already when treating $\tau =2n/N$ as a continuous parameter we took $%
N\rightarrow \infty $ limit implicitly. The remaining limit is $%
V_{k}\rightarrow 0$ and we know that $\Gamma (\mathbf{x},\mathbf{x}^{\prime
})\,\rightarrow \delta (\mathbf{x}-\mathbf{x}^{\prime })$ in this limit.
After setting $\Gamma (\mathbf{x},\mathbf{x}^{\prime })\,=\delta (\mathbf{x}-%
\mathbf{x}^{\prime })$ performing the $\mathbf{x}^{\prime }$ integral the
operator $\widehat{L}_{C}$ reduces to
\begin{equation}
\widehat{L}_{C}F(\mathbf{xv)}=\int \left[ F(\mathbf{xv}_{A},t)F(\mathbf{xv}%
_{B},t)-F(\mathbf{xv}_{C},t)F(\mathbf{xv},t)\right] \sigma (\mathbf{n,n}%
^{\prime })\,d^{3}\mathbf{u}\,d\mathbf{n}^{\prime }\,.  \label{m51}
\end{equation}
With this form of the $\widehat{L}_{C}$ the eq. (\ref{m40}) is the Boltzmann
equation.

Hence we have shown that in direct simulation algorithm for inhomogeneous
gas the one particle probability distribution satisfies the Boltzmann
equation. Now, how do we connect this to the Bird's NTC\ algorithm? Clearly
they are not the same. In fact our algorithm is not practical since great
majority of chosen pairs will not be in the same cell and therefore will not
make collisions.

In the time interval $\Delta t\,$we choose $\ \Delta n=RN^{2}\Delta
t/2\Omega $ pairs. The probability that each pair will make a collision
attempt in the $k^{th}$ cell is $p_{k}=(\Omega /V_{k})(N_{k}/N)^{2}.$ Let $%
n_{k}$ be the number of collision attempts that take place in $V_{k}$. The
expected value of $n_{k}$ is
\begin{equation}
\overline{n}_{k}=\Delta n\cdot p_{k}=\frac{RN_{k}^{2}}{2V_{k}}\Delta t.
\label{m60}
\end{equation}
This is the same as number of collision attempts in $V_{k}$ in Birds
algorithm. The difference is that in Birds algorithm the number of collision
attempts in each cell is fixed as $n_{k}=RN_{k}^{2}\Delta t/2V_{k}$ whereas
in our algorithm the $n_{k}$ has a probability distribution with a mean
value $RN_{k}^{2}\Delta t/2V_{k}$. The probability distribution for $n_{k}$
is given as
\begin{equation}
P(n_{k})=\frac{(\Delta n)!}{(\Delta n-n_{k})!\,(n_{k})!}%
(p_{k})^{n_{k}}(1-p_{k})^{\Delta n-n_{k}}.  \label{m71}
\end{equation}
In the limit of $V_{k}\rightarrow 0$ we have $p_{k}\rightarrow 0$ and the $%
P(n_{k})$ becomes the Poisson probability distribution
\begin{equation}
P(n_{k})=\frac{(\overline{n}_{k})^{n_{k}}}{(n_{k})!}\exp (-\overline{n}_{k}).
\label{m80}
\end{equation}
The width of distributions in eqs.(\ref{m71},\ref{m80}) is of order $\sqrt{%
\overline{n}_{k}}$. For large values of $\overline{n}_{k}$ we have $n_{k}/%
\overline{n}_{k}=1+O(1/\sqrt{\overline{n}_{k}})$ where $O(1/\sqrt{\overline{n%
}_{k}})$ is a term of order $1/\sqrt{\overline{n}_{k}}$.

Now we take the limit $N_{k}\rightarrow \infty $ and $O(1/\sqrt{\overline{n}%
_{k}})$ error term vanishes. In a more mathematical language, probability
that $n_{k}/\overline{n}_{k}=1$ is unity. Hence both methods approach each
other in the limit $N_{k}\rightarrow \infty $ and single particle
probability distribution in Bird's method too should satisfy the Boltzmann
equation (eq.(\ref{m40})) in this limit.

There is an important distinction in the limits taken for both method to
satisfy the Boltzmann equation. In our algorithm we take $N\rightarrow
\infty ,$ $\Delta t\rightarrow 0$ and $V_{k}\rightarrow 0$ limits. This does
not mean that number of particles in each cell ($N_{k}$) will go to
infinity. For example for a uniform density we have $N_{k}=(N/V)V_{k}.$ Here
$N\rightarrow \infty $ and $V_{k}\rightarrow 0$ limits does not imply
anything about $N_{k}$. $NV_{k}$ can remain finite and even can go to zero
and still our algorithm satisfies the Boltzmann equation. The Bird's
algorithm requires $N_{k}\rightarrow \infty $ to satisfy the Boltzmann
equation however and this is a more stringent requirement.

We did this analysis for the simplest case of one kind of gas without
internal degrees of freedom for clarity. It is very simple to generalize
this to the other cases by replacing the kernel $S$ in eq.(\ref{k50}) with $%
G_{pq}^{rs}$ in eq.(\ref{f20}) or with $K_{ij}^{\alpha \beta }$ in eq.(\ref
{g21}) or with $Q_{ij,pq}^{\alpha \beta ,rs}$ in eq.(\ref{h21}). Then the
Boltzmann equation will be replaced by the Wang Chang-Uhlenbeck equation but
all of the arguments will remain the same.

\section{Conclusions}

Let us list our contributions in this paper.

\begin{itemize}
\item  In this paper we introduced a general formalism for direct simulation
processes. We defined the direct simulation as a markov process with a
master equation and we found the master equation given in eq.(\ref{a51}).
Definition the DSMC\ algorithm as a stochastic process governed by a master
equation does not exist in the literature of the DSMC\ method to our
knowledge.

\item  Starting from the master equation we showed that the N-particle
probability density evolves towards microcanonical distribution as the
number of collisions go to infinity.

\item  We derived a hierarchy of equations similar to the BBGKY hierarchy
for the reduced probability densities given in eq.(\ref{a60})

\item  We showed that if AMC\ approximation is employed the single particle
probability distribution satisfies an equation given in eq.(\ref{a101}). In
the limit $N\rightarrow \infty $ this reduces to eq.(\ref{a130}) which is an
equation similar to the Boltzmann equation.

\item  We found the equations of the hierarchy in the limit $N\rightarrow
\infty $ (the eq.(\ref{a134}) )and showed that the ansatz $f^{(M)}(\mu
_{1},\mu _{2},...,\mu _{M};\tau )=f^{(1)}(\mu _{1}\mathbf{;}\tau
)\,f^{(1)}(\mu _{2}\mathbf{;}\tau )....f^{(1)}(\mu _{M}\mathbf{;}\tau )$
satisfies all the equations in the hierarchy provided the $f^{(1)}(\mu
\mathbf{;}\tau )$ satisfies the eq.(\ref{a130}). This ensures that in the
limit $N\rightarrow \infty $ the AMC is satisfied for all times if one
starts from an uncorrelated initial state.

\item  We gave two simple examples from direct simulation money games. The
discrete money game example has the nice feature that it is exactly solvable
and we observe from the solution that the approach to the equilibrium is
exponentially fast.

\item  We obtained the H-theorem and conservation of expectation values of
collision invariants. These results are familiar to most readers from the
standard treatments of the Boltzmann equation. But it is worth repeating
them here because although the equations are similar they are applied to
wide variety of different problems in the direct simulation setting, not
just to gases.

\item  We applied the formalism to the direct simulation Monte Carlo method
for real homogenous gases which is a standard method to solve the Boltzmann
equation. Introducing appropriate kernels we obtained NTC\ algorithm for a
homogenous gas and we showed that the appropriately normalized single
particle probability distribution satisfies Boltzmann equation for simple
homogenous gases and Wang Chang-Uhlenbeck equations for homogenous molecular
gases and their mixtures. The derivation of conservation of $\int f^{p}(%
\mathbf{v})\,d^{3}\mathbf{v}$ for mixture of gases without internal degrees
of freedom and $\sum_{i}\int f_{i}^{p}(\mathbf{v})\,d^{3}\mathbf{v}$ for
mixture of gases with internal degrees of freedom should be also familiar to
the reader from the standard treatments of the Boltzmann equation. The novel
feature of our derivation is the significant simplification that the
normalization of $T(\mu _{A},\mu _{B},\mu _{C},\mu _{D})$ given in the
equations (\ref{a30},\ref{f111},\ref{f111a},\ref{h120},\ref{h121}) provide
to obtain the result. If we try to obtain the same result from the Boltzmann
equation we would have to use the argument that the integrals in (\ref{f111},%
\ref{f111a},\ref{h120},\ref{h121}) are functions of the collision invariants.

\item  We introduced a new algorithm to do the DSMC\ calculations for an
inhomogeneous gas. Our algorithm is not practical for the actual practice of
the art because of wasting the great majority of the chosen pairs. We showed
that the single particle probability distribution satisfies the Boltzmann
equation in our algorithm in the limits $N\rightarrow \infty ,$ $\Delta
t\rightarrow 0$ and $V_{k}\rightarrow 0$. We also showed that Bird's
algorithm for DSMC converges to our algorithm if $N_{k}\rightarrow \infty $
is taken in addition to the limits $\Delta t\rightarrow 0$ and $%
V_{k}\rightarrow 0$. Birds algorithm requires more stringent requirements to
satisfy the Boltzmann equation. To prevent any misunderstanding we stress
here that our algorithm is not intended as a practical scheme to implement
DSMC calculations. The Bird's algorithm does not easily fit in the direct
simulation formalism presented in this paper whereas the algorithm we
presented does. We showed that our algorithm gives the Boltzmann equation in
the limits $N\rightarrow \infty ,$ $\Delta t\rightarrow 0$ and $%
V_{k}\rightarrow 0$ and we also showed that our algorithm and Bird's
algorithm converges to each other if we go to more stringent limit of $%
N_{k}\rightarrow \infty $. Therefore we proved indirectly that Birds
algorithm satisfies Boltzmann equation in the limit $N_{k}\rightarrow \infty
$, $\Delta t\rightarrow 0$ and $V_{k}\rightarrow 0$. Therefore we introduced
our algorithm as a tool to study convergence of Bird's method and not as a
practical way of doing DSMC calculations.
\end{itemize}

Meaning of the convergence here should be interpreted according to the
ensemble theory of statistical mechanics. We imagine practically infinite
number of identical systems (computers with human operators) doing the same
direct simulation and call this the ensemble. The $f^{(1)}(\mu ;\tau )d\mu $
represents ratio of number of particles in $d\mu $ to the total number of
particles averaged over all the ensemble. When you perform a direct
simulation on a computer you are just one member of the ensemble. Your
results will show statistical fluctuations. But when you do the same
simulation many times with different initial states chosen according to a
uncorrelated probability distribution $f^{(N)}(\mu _{1},\mu _{2},...,\mu
_{N};n=0)=\,h(\mu _{1})h(\mu _{2})....h(\mu _{N})$ you form your own
ensemble and averages over them will nicely follow $f^{(1)}(\mu ;\tau )$
obtained by solving eq.(\ref{a130}) with the initial value $f^{(1)}(\mu
;\tau =0)=h(\mu ) $.

This work can generalize to chemical reactions and radiative processes in a
more or less straightforward fashion. But there are enough number of
subtleties such that we leave them to future publications.

A simplified version of this paper\cite{AJP} containing only one kind of
homogenous gas without internal degrees of freedom is published in American
Journal of Physics. The material in that paper makes a small fraction of the
material in this paper. The present paper contains much new material and
overlap between the two papers is small.

\end{document}